\documentclass[twocolumn, 10pt,journal]{IEEEtran}
\ifCLASSINFOpdf
\else
\fi
\usepackage{mathrsfs}
\usepackage{graphicx}
\usepackage{epsfig}
\usepackage{amsmath}
\usepackage{bm}
\usepackage{float}
\usepackage{multirow}
\usepackage{color}
\usepackage{subfig}
\usepackage{amsmath}
\usepackage{algorithm}
\usepackage{algorithmic}
\usepackage{cite}
\usepackage{amsthm}
\usepackage{amssymb}
\usepackage{tabularx}
\usepackage{booktabs}
\usepackage{color}

\newenvironment{sequation*}{\begin{equation*}\small}{\end{equation*}}

\theoremstyle{definition}

\theoremstyle{remark}

\theoremstyle{proposition}

\long\def\symbolfootnote[#1]#2{\begingroup%
\def\thefootnote{\fnsymbol{footnote}}\footnote[#1]{#2}\endgroup}

\ifCLASSOPTIONcompsoc
\usepackage[nocompress]{cite}
\else
\usepackage{cite}
\fi

\begin{document}

\bibliographystyle{IEEEtran}

\title{Distributed Virtual Resource Allocation in Small Cell Networks with Full Duplex Self-backhauls and Virtualization}

\author{Lei Chen, F.~Richard~Yu, \textit{Senior Member, IEEE}, Hong Ji, \textit{Senior Member, IEEE}, Gang Liu, and Victor C.M. Leung, \textit{Fellow, IEEE}\\
\thanks{Copyright (c) 2015 IEEE. Personal use of this material is permitted. However, permission to use this material for any other purposes must be obtained from the IEEE by sending a request to pubs-permissions@ieee.org.}
\thanks{This work is jointly supported by Project 61271182 and 61302080 of the National Natural Science Foundation of China.
}
\thanks{L. Chen, H. Ji and G. Liu are with the Key Laboratory of Universal Wireless Communications, Ministry of Education, Beijing University of Posts and Telecommunications,
Beijing, P.R. China (e-mail:chenlei1989bupt@gmail.com, jihong@bupt.edu.cn and buptgangliu@gmail.com).}
\thanks{F. R. Yu is with the Dept. of Systems and Computer Eng., Carleton University, Ottawa, ON, Canada (e-mail: Richard.yu@carleton.ca).}
\thanks{V. C. M. Leung is with the Department of Electrical and Computer Engineering, the University of British Columbia, Vancouver, BC V6T 1Z4 Canada
(e-mail: vleung@ece.ubc.ca).}
}
\maketitle

\begin{abstract}
Wireless network virtualization has attracted great attentions from both academia and industry. Another emerging technology for next generation wireless networks is in-band full duplex (FD) communications. Due to its promising performance, FD communication has been considered as an effective way to achieve self-backhauls for small cells. In this paper, we introduce wireless virtualization into small cell networks, and propose a virtualized small cell network architecture with FD self-backhauls. We formulate the virtual resource allocation problem in virtualized small cell networks with FD self-backhauls as an optimization problem. Since the formulated problem is a mixed combinatorial and non-convex optimization problem, its computational complexity is high. Moreover, the centralized scheme may suffer from signaling overhead, outdated dynamics information, and scalability issues. To solve it efficiently, we divide the original problem into two subproblems. For the first subproblem, we transfer it to a convex optimization problem, and then solve it by an efficient alternating direction method of multipliers (ADMM)-based distributed algorithm. The second subproblem is a convex problem, which  can be solved by each infrastructure provider. Extensive simulations are conducted with different system configurations to show the effectiveness of the proposed scheme.
\end{abstract}


\section{Introduction}

By abstracting and sharing resources among different parties, \textit{virtualization} can significantly reduce the cost of equipment and management in networks \cite{chowdhury2010survey}. With the tremendous growth in wireless traffic and service, it is inevitable to extend virtualization to wireless networks \cite{LY15,LYZ15}. In wireless virtualization, physical wireless network infrastructure and physical radio resources are abstracted and sliced into virtual wireless resources, which can be shared by multiple parties. After virtualization, the wireless network infrastructure owned by an infrastructure provider (InP) can be decoupled from the services that it provides. At the same time, mobile virtual network operators (MVNOs) provide services to users. Since the physical resources are abstracted and sliced into virtual resources, it is possible that different MVNOs coexist on the same InP to share the infrastructure and radio spectrum resources, which enables the reducing of capital expenses (CapEx) and operation expenses (OpEx)\cite{LY15}.

The authors of \cite{HS11} discussed the control and management frameworks of wireless network virtualization. Virtual resource sharing mechanisms were investigated in \cite{FK13}, where the dynamic interactions among MVNOs and InPs are modeled as a stochastic game. In addition, there are some other works focusing on the virtualization of certain specific wireless networks. For example, in the context of cellular networks, Zaki \textit{et al.} \cite{ZZGT11} proposed a virtualization framework for long-term evolution (LTE) systems, in which a supervisor is used to virtualize the eNB and manage the physical resources. For wireless local area networks (WLANs) virtualization, a SplitAP architecture was proposed in \cite{BVSR10}, and a resource sharing algorithm based on control theory was designed in \cite{BSPN12}. Moreover, virtualization techniques for WiMAX networks \cite{KMZR12} and wireless-optical heterogeneous networks \cite{TAZRNS13} were also studied.

Although some excellent researches have been done for wireless virtualization, most existing works do not consider \textit{small cell networks with self-backhauls}. Recently, small cell networks have been regarded as one of the key components of next generation cellular networks to improve spectrum efficiency and energy efficiency \cite{Haijun2015R,BY14}. Traditionally, there are two kinds of backhauls in small cell networks: wired backhaul (e.g., optical \cite{RIOR2013} or DSL \cite{andrews2012femtocells}) and wireless backhaul (e.g.,  microwave \cite{WCXR2014} or millimeter waves \cite{STDJT2013}). Since these traditional backhauls are very expensive for infrastructure deployments, self-backhauled small cells have attracted great attentions from both academia and industry \cite{SPQT2014,BIB15}. Wireless self-backhauling can improve reachability and coverage by easing connectivity between nodes \cite{BIB15}. In \cite{SPQT2014}, a self-backhauling scheme was proposed, in which the self-backhaul link uses the same spectrum with the small cell downlink, but on different time slots.

Another emerging technology for next generation wireless networks is in-band \textit{full duplex} (FD) communications. With the recent advances of self-interference cancellation techniques \cite{SSGBRW13,LiuGang2014}, it is possible for radios to transmit and receive simultaneously in the same frequency band. Due to its promising performance, FD communication has been considered as an effective way to achieve self-backhauls for small cells. The authors of \cite{HBJJ2014} and \cite{LiuGang2014} made fine attempts in this direction. Nevertheless, no detail has been reported.

Despite the potential vision of small cell networks with FD self-backhauls and virtualization, many research challenges remain to be addressed. One of the main research challenges is \textit{resource allocation}, which plays an important role in traditional wireless networks \cite{FR2012,AHSWN_DWD15,XYJL12}. When wireless virtualization and FD self-backhauls are jointly considered, the problem of resource allocation becomes even more challenging, since the backhaul and access links are coupled, which depend on virtual resource allocation and self-interference cancellation performance. To the best of our knowledge, the problem of virtual resource allocation in small cell networks with FD self-backhauls and virtualization has not been studied in previous works. The distinct features of this paper are summarized as follows.

\begin{itemize}
\item We introduce wireless virtualization into small cell networks, and propose a virtualized small cell network architecture, where multiple InPs and multiple MVNOs coexist. In the proposed virtualized small cell networks, in addition to radio spectrum, both macro base stations (MBSs) and small cell base Stations (SBSs) from different InPs are virtualized as virtual resources, which can be dynamically shared by users from different MVNOs.
\item We formulate the virtual resource allocation problem in virtualized small cell networks with FD self-backhauls as an optimization problem, which maximizes the total utility of all MVNOs, considering not only the revenue earned by serving users but also the cost paid to InPs for consuming power, spectrum, and backhaul resources. In addition, we take the residual self-interference of FD communications into account in the formulated problem.
\item Since the formulated problem is a mixed combinatorial and non-convex optimization problem, its computational complexity is high. Moreover, the centralized scheme may suffer from signaling overhead, outdated dynamics information, and scalability issues. To solve it efficiently, we divide the original problem into two subproblems. For the first subproblem, we transfer it to a convex optimization problem, and then solve it by an efficient \textit{alternating direction method of multipliers}\cite{BPCPE11} (ADMM)-based distributed algorithm, in which the InPs and virtual resource manager (VRM) only need to solve their own problems without exchange of channel state information with fast convergence rate. The second subproblem is a convex problem, which  can be solved by each InP. 
\item Extensive simulations are conducted with different system configurations to verify the effectiveness of the proposed scheme. It's shown that we can take the advantages of both wireless network virtualization and FD self-backhauls with the proposed distributed virtual resource allocation algorithm. InPs, MVNOs, and users can benefit from the proposed resource allocation scheme.
\end{itemize}

The rest of paper is organized as follows. The proposed virtualized small cell networks architecture and the FD self-backhaul mechanism are described in \ref{Sec 2: Network model}. The resource allocation problem is formulated in \ref{Sec 3: problem formulation}. Then we divide the formulated problem into two subproblems and the solution details are described in \ref{Sec 4: solution}. Simulation results are discussed in \ref{Sec 5: simulation}. Finally, we conclude this study in Section \ref{Sec 6: conclusion}.

\section{System Model}
\label{Sec 2: Network model}
In this section, we first describe the virtualized small cell network architecture. Then we present the FD self-backhauling mechanism where the SBSs can transmit and receive data on the same spectrum simultaneously.

\subsection{Virtualized Small Cell Network Architecture}
\begin{figure}[!t]
\centering
\includegraphics[width=0.48\textwidth]{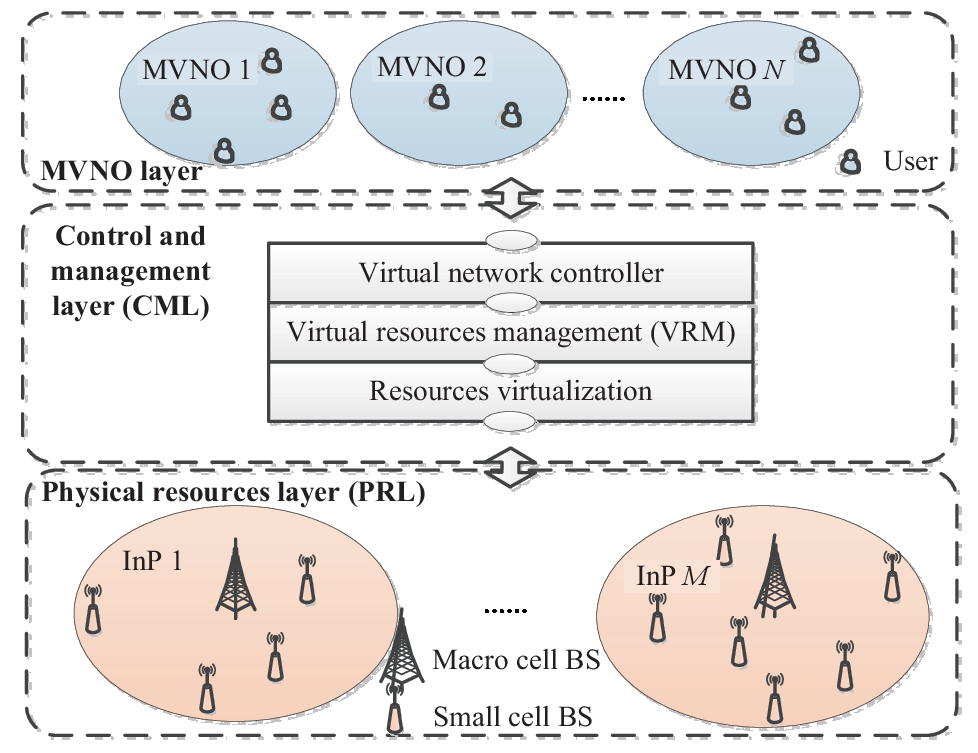}

\caption{A virtualized small cell network architecture.}
\label{fig: virtualization_arc}
\end{figure}

We present a virtualized small cell network architecture with multiple InPs and multiple MVNOs, as shown in Fig. \ref{fig: virtualization_arc}. There are  $M$ InPs offering wireless access services in a certain geographical area. Each InP deploys and manages a cellular network with one MBS and several self-backhauled SBSs. There are $N$ MVNOs, which provide various services to their subscribers through the same substrate networks. Following the general frameworks of wireless network virtualization \cite{FK13,LY15m}, the virtualized small cell network architecture consists of three layers: the \textit{physical resource layer} (PRL), the \textit{control and management layer} (CML), and the \textit{MVNO layer}. The PRL, including base stations (BSs), spectrum, power and backhauls from different InPs, is responsible for providing available physical resources. Moreover, the PRL also provides CML with the interfaces needed to control resources. The CML virtualizes the physical resources from different InPs and enables the sharing for MVNOs.
Then, the CML manages and allocates the virtual resources to different MVNO users. The resource management functions in CML are realized by a \textit{virtual network controller} and a \textit{virtual resource manager} (VRM). The virtual network controller of MVNOs is responsible to collect the resource consumption prices negotiated with InPs, and the users' information (e.g., payment information and QoS requirements) from MVNOs, then feedback the resource allocation results to MVNOs for the purpose of finishing the settlement between MVNOs and InPs. To maximize the total utility of all MVNOs, the VRM is responsible to dynamically allocate the virtual resources from multiple InPs to different MVNO users. Through the virtualization architecture above, each MVNO can have a virtual network composed of the substrate networks from multiple InPs. Hence each user can get services via different access points (either MBSs or SBSs) from different InPs.

We assume that the spectrum bandwidth of the $m$-th InP is $B_m$. The transmit power of the MBS and the transmit power of the SBSs, are $P_m$ and $P_m^s$, respectively. $\bm{S}_m$ is used to represent the set of SBSs that belong to the $m$-th InP. Let $S_m^j$ be the $j$-th SBS in $\bm S_m$. For ease of presentation, we use $j\in \bm S_m$ to represent $S_m^j\in\bm S_m$ in this paper. The set of users of MVNO $i$ is denoted as $\bm{U}_i$. Then, let $\bm{U}=\cup_{i=1}^N\bm{U}_i$ be the set of all the users and $\|{\bm{U}}\|$ be the total number of users. For the users, there are two access choices: MBS or self-backhauled SBS. Those users who are associated to the $m$-th MBS is denoted by $\bm U_m$. Similarly, those users who access to the $j$-th SBS in the $m$-th InP is denoted by $\bm{U}_m^{s_j}$.  To facilitate the formulation of the virtual resource allocation problem, the self-backhauling mechanism via in-band FD communications will be introduced in the next subsection.

\subsection{Small Cell Self-backhauling Mechanism Based on Full Duplex Communications}

\begin{figure}[!t]
\centering
\includegraphics[width=0.48\textwidth]{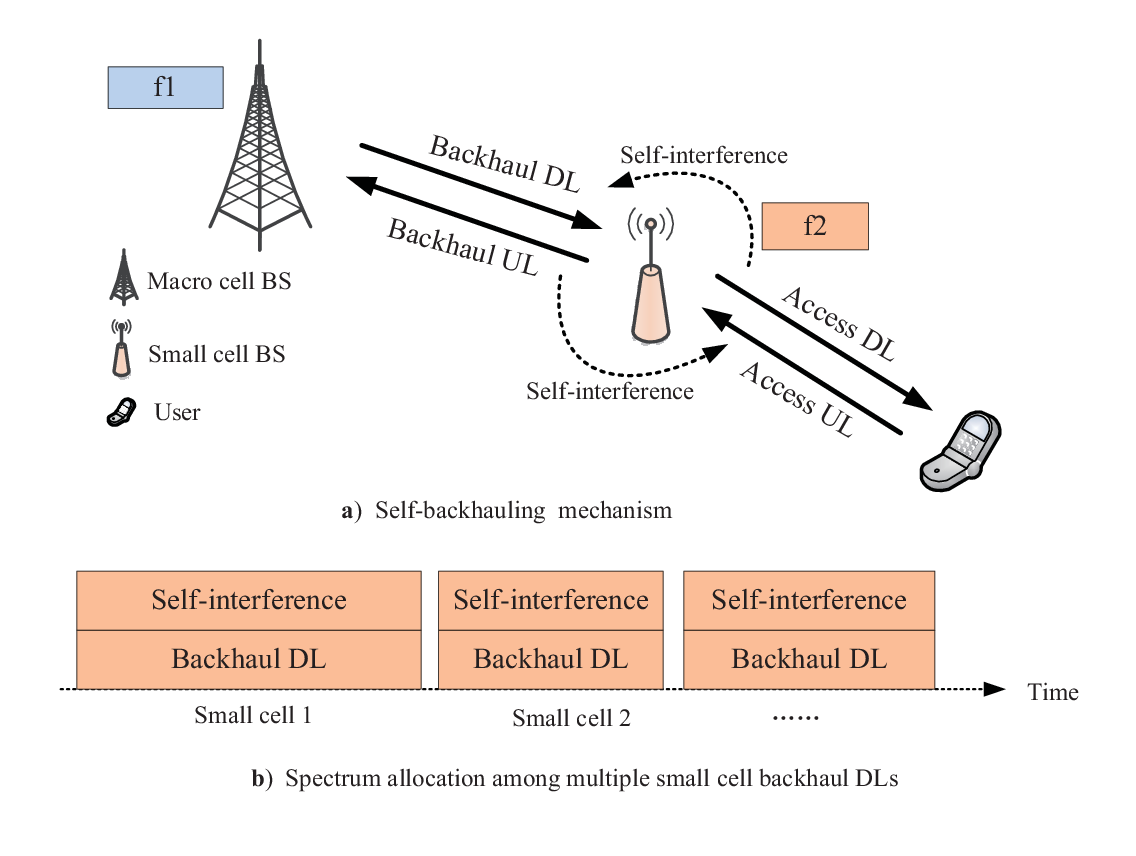}
\caption{A small cell self-backhauling mechanism based on full duplex communications.}
\label{fig: self-backhaul mechanism}
\end{figure}

As shown in Fig. \ref{fig: self-backhaul mechanism}(a), SBSs are equipped with FD hardware, which enables them to backhaul data for themselves. In the downlink (DL), a SBS can receive data from the MBS, while simultaneously transmitting to its users on the same frequency band. In the uplink (UL), a SBS can receive data from the users, while simultaneously transmitting data to the MBS on the same frequency band. In this mechanism, the SBS can effectively backhaul itself, eliminating the need for a separate backhaul solution and a separate frequency band. Therefore, self-backhauling can significantly reduce the cost and complexity of rolling out small cell networks. In order to distinguish DL from UL in access and backhaul transmissions, we call the relevant links as access UL, access DL, backhaul UL, and backhaul DL, respectively. Due to the limitation of self-interference cancellation technologies, the backhaul DL and access UL will suffer some self-interference from access DL and backhaul UL, respectively. Different from the FD relay mechanism in \cite{LiuGang2014}, the spectrum can be reused by different SBSs and the SBSs can allocate resource to their users flexibly in our self-backhaul scheme. Compared to DL, UL usually has less traffic. If the transmission of DL is satisfied, the transmission of UL will also be satisfied. As a result, we focus on the transmission of DL in this paper.

In this paper, the orthogonal spectrum reuse pattern is adopted, where the spectrum is divided for the MBS and SBSs to avoid the inter-layer interference \cite{XAG2013}. As shown in Fig. \ref{fig: self-backhaul mechanism}(a), we divide the spectrum of InP $m$ into two parts: $\alpha_m B_m$ as $\mathit{f_1}$ for the MBS and $(1-\alpha_m)B_m$ as $\mathit{f_2}$ for the SBSs. In DL, the MBS  transmits data to not only macro users on $\mathit{f_1}$ but also SBSs on $\mathit{f_2}$. Similarly, the MBS receives the data from macro users on $\mathit{f_1}$ and SBSs on $\mathit{f_2}$ in UL. At the same time, SBSs transmit and receive data to their users on $\mathit{f_2}$. Obviously, the spectrum indicator vector $\bm\alpha = (\alpha_1, \alpha_2, ..., \alpha_M)$, which decides the throughput of backhaul and access link, plays an important role in achieving small cell self-backhauls.

For user $u\in\bm{U}_i$, the VRM will decide its association BS and allocate some resources to it from the BS. We denote by $x_u^{m,j}$ and $y_u^{m,j}$ the user's association indicator and allocated resources ratio (Note that this resources mean time slot resources), respectively. When a user is associated with one BS, $x_u^{m,j}=1$, otherwise $x_u^{m,j}=0$. Further, $j=0$ means the user is associated to the $m$-th MBS, $j\neq0$ means the user is associated to $S_m^j$. Similarly, $y_u^{m,0}$ denotes the resource ratio that the user gets from the $m$-th MBS, and $y_u^{m,j} (j\neq 0)$ denotes the resource ratio that the user gets from $S_m^j$. Obviously, $y_u^{m,j}$ is related to $x_u^{m,j}$. Only when $x_u^{m,j}=1$, $y_u^{m,j}$ will be meaningful.

For the access DL, we denote by $R_u^{m,j}$ the achievable link rate of one user. Generally, $R_u^{m,j}$ and $R_m^{s_j}$ are logarithmic functions of signal to interference and noise ratio (SINR). For macro users, they do not suffer interference from other BSs because of the different spectrum bands among InPs and the OSR pattern between macro and SBSs, so the achievable link rate can be expressed as
\begin{align}
&R_u^{m,0}
= \alpha_mB_m\log { \left\{1+\frac{P_mh_u^{m,0}}{\sigma^2}\right\} } \label {rate-u-M}.
\end{align}
For small cell users, they suffer co-channel interference from other SBSs in the same InP, so the achievable link rate can be expressed as
\begin{align}
&R_u^{m,j}=\notag \\
&(1-\alpha_m)B_m\log{\left\{1+\frac{P_m^sh_u^{m,j}}{\sum\limits_{k\neq j,k\in{\bm{S_m}}}P_m^sh_u^{m,k}+\sigma^2}\right\}}, j\neq0. \label {rate-u-S}
\end{align}
In our virtualized small cell networks, each user from any MVNO can access to either a MBS or a SBS in any InP. Hence, for a given user $u$, we define $C_u$ as its overall long-term rate, which can be expressed as follows.
\begin{equation}
\label{capacity-u-m-s}
C_u^m=x_u^{m,0}\underbrace{y_u^{m,0}R_u^{m,0}(\alpha_m)}_{C_u^{m,0}}+\sum\limits_{j\in \bm{S}_m}x_u^{m,j}\underbrace{y_u^{m,j}R_u^{m,j}(\alpha_m)}_{C_u^{m,j}},
\end{equation}
where $C_u^{m,0}$ and $C_u^{m,j}$ are the overall long-term rates of user $u$ getting from the $m$-th MBS and $S_m^j$, respectively.
For the backhaul DL, the MBS transmits data to SBSs on $\mathit{f_2}$, and the SBSs buffer these data to transmit to small cell users. We define the backhaul link rate of $S_m^j$ by $R_m^{s_j}$. When the SBS receives data from the MBS, it transmits data to its users at the same time, which results in the self-interference (SI). The value of SI is determined by self-interference cancellation technologies, and is proportional to the transmission power of SBS DL, which could be expressed as $\textmd{SI}=\vartheta_mP_m^s$. $\vartheta_m$ represents the residual self-interference gain.  Different self-interference cancellation technologies may result in different $\vartheta_m$. Any self-interference cancellation technology  can be applied at the SBSs, and the analysis in this paper is a general case. In addition, a SBS also suffers co-channel interference from other SBSs because they use the same spectrum $\mathit{f_2}$, so the achievable link rate can be expressed as
\begin{align}
&R_m^{s_j}=\notag\\
&(1-\alpha_m)B_m\log{\left\{1+\frac{P_mh_m^{s_j}}{\vartheta_mP_m^s+\sum\limits_{k\neq j, k\in{\bm{S_m}}}P_m^sh_{s_j}^{s_k}+\sigma^2}\right\}}. \label {rate-s-m}
\end{align}
In each InP, SBSs share the spectrum $\mathit f_2$, but the spectrum resource must be divided in time domain or frequency domain (TDD or FDD) to avoid interference. If FDD is applied, the SBS will have less spectrum receiving data than that transmitting data, which increases the difficulty to investigate SI. As a result, TDD model is adopted in this paper, as shown in Fig. \ref{fig: self-backhaul mechanism}(b). We use $z_m^{s_j}$ to denote the time slot ratio occupied by SBS $S_m^j$. If we denote by $C_m^{s_j}$ the overall long-term rate of backhaul DL from the $m$-th MBS to $S_m^j$, it can be expressed as $C_m^{s_j}=z_m^{s_j}R_m^{s_j}$.

In (\ref{rate-u-M}), (\ref{rate-u-S}), and (\ref{rate-s-m}), $\sigma^2$ denotes the noise power level, $h_u^{m,j}$ denotes the channel gain between one user and macro (small) cell BS and $h_m^{s_j}$ denotes the channel gain between MBS and SBS. In general, the channel gain includes path loss, shadowing and antenna gain.  The association is assumed to be carried out in a large time scale compared to the dynamics of channels. The SINR for association is averaged over the association time, and thus it is a constant regardless of the dynamics of channels (i.e., fast fading is averaged out). As for resource allocation, we assume that resource allocation is carried out well during the channel coherence time, and thus the channel can be regarded as static during each resource allocation period. This model is applicable for low mobility environment.

The notations used in this paper are presented in Table \ref{table_1}.
\begin{table}[!t]
\centering
\caption{Notation}
\label{table_1}
\begin{tabular}{ll}
\toprule
Notation & Definition\\
\midrule
${\bm{U}}$    & set of users\\
${\bm{S}}$    & set of SBSs\\
$\bm\alpha$   & spectrum allocation indicator vector\\
$M$           & the number of InPs\\
$N$           & the number of MVNOs \\
$m$           & index of MBSs\\
$i$           & index of MVNOs \\
$j$           & index of SBSs\\
$P_m$         & transmit power spectrum density of macro BS of InP $m$\\
$P_m^s$       & transmit power spectrum density of SBSs of InP $m$ \\
$B_m$         & the bandwidth of the $m$-th InP' spectrum \\
$h_m^{s_j}$   & channel gain between $S_m^j$ to the $m$-th macro BS\\
$h_u^{m, j}$  & channel gain between user $u$ to the $j$-th SBS in InP $m$ \\
$R_u^{m, j}$  & achievable link rate between user $u$ to BSs of InP $m$ \\
$R_m^{s_j}$   & achievable backhaul link rate of the $j$-th SBS in InP $m$\\
$x_u^{m, j}$  & user cell association indicator, \\
$y_u^{m, j}$  & user resource allocation  indicator \\
$z_m^{s_j}$   & SBS resource allocation indicator\\
$C_u^{m, j}$  & rate of user $u$ getting from BSs of $m$-th InP\\
$C_u^m$       & sum rate of user $u$ getting from all BSs of $m$-th InP\\
$C_m^{s_j}$   & rate of the backhaul link of the $j$-th SBS in InP $m$\\
\bottomrule
\end{tabular}
\end{table}

\section{Problem formulation}
\label{Sec 3: problem formulation}
In this section, we formulate the virtual resource allocation problem in virtualized small cell networks with FD self-backhauls. Firstly, we present the utility functions for users, MVNOs, and InPs. Then, the problem of virtual resource allocation is formulated to maximize the total utility of all MVNOs in a centralized manner at the VRM.
\subsection{Utility Function Definition}
In our virtualized small cell networks, users pay to their MVNO for getting services. 
Hence, for a given user $u$, the utility function can be defined as the difference between her/his service rate and the money she/he paid, which is expressed as
\begin{equation}
\label{utility_user}
G_u=\sum\limits_{m=1}^M\varsigma C_u^m-\delta_u,
\end{equation}
where $\varsigma$ means the profit of users for per bit rate. For simplicity, we assume $\varsigma=1.$
MVNOs purchase resources (including spectrum,time slot, power and backhaul) from InPs to provide services to their users. The responsibility of the VRM is to efficiently allocate the virtual resources to maximize the total utility of all MVNOs. In this paper, we consider a long-term rate-aware utility function for MVNOs, which is defined as the difference between the total income of all MVNOs earned by serving the users and the total resource consumption cost.
For the $i$-th MVNO, its utility can be expressed as:
\begin{align}
\label{G:MVNO}
G_{MVNO}=\sum\limits_{m=1}^M\sum\limits_{u\in\bm U_i}\delta_uC_u^m-\sum\limits_{m=1}^M\gamma^mT_i^m-\sum\limits_{m=1}^MQ_i^m.
\end{align}
The first term represents the income of the MVNO from providing services to users, and $\delta_u$ represents the money user $u$ has paid to its subscribed MVNO. The second term defines the cost of MVNO for using the spectrum and power resource, and $\gamma^m$ represents the deliberated price between the $m$-th InP and MVNOs. $T_i^m$ can be expressed as
\begin{align}
T_i^m=&\sum\limits_{u\in\bm U_m \& u\in\bm U_i}\left\{x_u^{m, 0}y_u^{m, 0}(\alpha_mB_m\cdot P_m)\right.\notag\\
&\left.+w_m\sum\limits_{j\in\bm S_m}x_u^{m, j}y_u^{m, j}\left\{(1-\alpha_m)B_m\cdot P_m^s\right\}\right\},
\end{align}
where \textit{bandwidth-power product} is used to quantify the consumed wireless resources of the access cell \cite{TLAA2013}, coefficient $w_m$ specifies the weight of small cell resources with respect to the MBS resources. As is well known, the load unbalance problem is a serious problem in small cell networks, which leads to low utilization of SBSs\cite{QBYA2013},\cite{FR2012}. As a result, we choose $w_m<1$ to attract more users to SBSs. This method  was also used in \cite{SHJ2013}. The last term represents the cost of MVNO for using the backhaul resource of InPs, which depends on the amount of backhaul data of MVNO and the type of backhaul technique. It's obvious that the amount of backhaul data for the  $i$-th MVNO in $S_m^j$ can be expressed as $\sum\limits_{u\in\bm U_i\&u\in\bm U_m^{s_j}}x_u^{m, j}y_u^{m, j}R_u^{m,j}$. We define the price of different backhaul techniques for backhaul \textit{one bit} by the price function $g(\eta_m)$, where $\eta_m$ represents the backhaul technique type of the $m$-th InP (e.g., optical, micowave, DSL, cell communication backhaul). Thus, the cost of MVNO for using backhaul resources can be expressed as
\begin{equation}
Q_i^m=\sum\limits_{j\in\bm S_m}g(\eta_m)\sum\limits_{u\in\bm U_i\&u\in\bm U_m^{s_j}}x_u^{m, j}y_u^{m, j}R_u^{m,j}.
\end{equation}
 Actually, the FD self-backhaul just utilizes part of MBS power on SBS access spectrum to achieve backhaul rather than additional spectrum or infrastructure, which is cheaper than traditional backhaul methods. Mathematically, $g(\eta_m)$ of our proposed FD self-backhaul is expressed as $g(\eta_m)=(1-\alpha_m)P_m z_m^{s_j}$.

As mentioned earlier, the responsibility of the VRM is to efficiently allocate the virtual resources for the purpose of maximizing the total utility of all MVNOs. In order to guarantee the fairness during the resource allocation process, we change the utility of MVNO as follows by adopting the method in \cite{QBYA2013},
\begin{align}
G_{MVNO}'=\sum\limits_{m=1}^M\sum\limits_{u\in\bm U_i}\delta_u\mathbb{U}(C_u^m)-\sum\limits_{m=1}^M\gamma^mT_i^m-\sum\limits_{m=1}^MQ_i^m ,
\end{align}
where function $\mathbb{U}(C_u^m)$ is defined as
\begin{equation}
\mathbb{U}(C_u^m)=x_u^{m, 0}\log\{y_u^{m, 0}R_u^{m, 0}\}+\sum\limits_{j\in\bm S_m}x_u^{m, j}\log\{y_u^{m, 0}R_u^{m, j}\}.
\end{equation}
Then the utility function of \textit{VRM} will be changed as
\begin{align}
\label{G:VRM_final}
&G_{VRM}'=\sum\limits_{i=1}^NG_{MVNO}'\notag \\
&=\sum\limits_{i=1}^N\left\{\sum\limits_{m=1}^M\sum\limits_{u\in\bm U_i}\delta_u\mathbb{U}(C_u^m)-\sum\limits_{m=1}^M\gamma^mT_i^m-\sum\limits_{m=1}^MQ_i^m\right\}.
\end{align}

For an InP, it not only leases its spectrum and power resource to MVNOs and earn money from them, but also rents or deploys infrastructure to backhaul data. So the utility function of the $m$-th InP can be expressed as
\begin{equation}
\label{Utility_InP}
G_{InP}=\gamma^m\sum\limits_{i=1}^NT_i^m+\sum\limits_{i=1}^NQ_i^m-\mathcal{O},
\end{equation}
where $\mathcal{O}$ means the cost of renting or deploying backhaul infrastructure. For simplicity, we assume that all the backhaul income from MVNOs  is used to pay for the rent or deployment of backhaul infrastructure in non-self-backhauled small cell networks, which means $\sum\limits_{i=1}^NQ_i^m=\mathcal{O}$.



\subsection{Virtual Resource Allocation Problem Formulation}
In each resource allocation cycle, the VRM needs to dynamically allocate the virtual resources, including MBSs, SBSs, and spectrum, to MVNOs. The objective of this paper is to develop a slot-by-slot virtual resource allocation algorithm that maximizes the total utility in (\ref{G:VRM_final}) under the following constraints to determine $\bm{\alpha}$, $\bm X$, $\bm Y$ and $\bm Z$.

Firstly, in our proposed small cell self-backhaul mechanism, the DL data of small cell users are transmitted from MBS to SBSs firstly, and then the SBSs store and forward it to their users. In this process, the throughput of backhaul DL decides the throughput of access DL. Hence, we have the throughput constraint of SBSs
\begin{equation}
C1:  \sum\limits_{u\in \bm{U}_m^{s_j}}x_u^{m,j}y_u^{m,j}R_u^{m,j}\leq z_m^{s_j}R_m^{s_j} ~ j\neq 0. \\
\end{equation}


Secondly, for simplicity, each user can only be served either by one MBS or by one SBS. Therefore, the cell association indicator should also satisfy the following constraints.
\begin{align}
& C2: x_u^{m,j}\in\{0,1\}, \\
& C3: \sum\limits_{m=1}^M\sum\limits_{j=0 or j\in \bm S_m} x_u^{m,j}\leq 1.
\end{align}

Thirdly, one BS schedules its associated users on time dimension and $y_u^{m,j}$ represents the schedule ratio of user $u$, the SBSs in the same InP share the frequency $f_2$ on time dimension to access and backhaul. $z_m^{s_j}$ defines the schedule ratio of SBS $S_m^j$ to backhaul. So, the following condition must be satisfied.
\begin{align}
& C4: 0\leq y_u^{m,j} \leq 1 ~  \forall m, j ,\\
& C5: \sum\limits_{u\in\bm U_m^0 or \bm U_m^{s_j}} x_u^{m,j}y_u^{m,j}\leq 1,  \\
& C6: 0\leq z_m^{s_j}\leq1 ~ \forall m,j, \\
& C7: \sum\limits_{j\in\bm S_m}z_m^{s_j}\leq1 ~ \forall m.
\end{align}

Furthermore, the spectrum allocation indicator $\alpha_m$ will take a value in the  interval of $[0,1]$. Mathematically, it can be expressed as follows.
\begin{align}
& C8: 0\leq \alpha_m \leq 1.
\end{align}

Then, based on the above constraints, the problem of resource allocation can be formulated as follows
\begin{align}
\mathscr{P}:&\max\limits_{\bm X, \bm Y, \bm Z, \bm\alpha}~ G_{VRM}'\\
& s.t. ~ C1, C2, C3, C4, C5, C6, C7, C8.\notag
\end{align}


\section{Distributed Virtual Resource Allocation Algorithms}
\label{Sec 4: solution}
It can be observed that the considered problem is combinatorial and non-convexity. The combinatorial nature comes from the integer constraint $C2$. The non-convexity is caused by the objective function and constraint $C1$. As a result, a brute force approach can be used  to obtain the optimal virtual resource allocation policy. However, such a method is computationally infeasible for a large system and does not provide useful system design insight. To reduce the computational complexity,
firstly, we assume the spectrum allocation indicator vector $\bm\alpha$ is fixed, and then we convert the original problem into a convex problem by variable transformation and perspective function theory to solve $\bm X$, $\bm Y$ and $\bm Z$. Secondly, based on the obtained results, we can prove that the problem is convex problem about $\bm \alpha$, then it's easy to get the optimized $\bm\alpha$ by some convex optimization algorithms. Furthermore, we come back to first step with the result of $\bm\alpha$ to get the new values of $\bm X$, $\bm Y$, and $\bm Z$. By iterations like this for a number of times, the values of  $\bm X$, $\bm Y$, $\bm Z$ and $\bm\alpha$ will converge, which can be seen as the solution of original problem $\mathscr{P}$. In Subsection \ref{section 4-A}, we solve $\bm X$, $\bm Y$ and $\bm Z$ by assuming $\bm\alpha$ is given. Subsection \ref{section 4_B} will introduce how to solve $\bm\alpha$ when $\bm X$, $\bm Y$ and $\bm Z$ are fixed. In Subsection \ref{section 4-C}, we describe the whole process of solution and the convergence of the proposed algorithms.

\subsection{Solving $\bm{X}$, $\bm Y$ and $\bm Z$ under Fixed Spectrum Allocation Indicators $\bm{\alpha}$} \label{section 4-A}
For any given spectrum allocation indicator vector $\bm\alpha$, the original problem reduces to the following problem:
\begin{align}
\mathscr{\bm {P}}_1:~ &\max\limits_{\bm X, \bm Y,\bm Z}~\sum\limits_{i=1}^N\left\{\sum\limits_{m=1}^M\sum\limits_{u\in\bm U_i}\delta_u\mathbb{U}(C_u^m(\bm X, \bm Y))\right.\notag \\
&\left.-\sum\limits_{m=1}^M\gamma^mT_i^m(\bm X, \bm Y)-\sum\limits_{m=1}^MQ_i^m(\bm X, \bm Y, \bm Z)\right\} \\
s.t. ~ & \ddot{C1}: \sum\limits_{u\in\bm U_m^{s_j}}x_u^{m, j}y_u^{m, j}R_u^{m, j}\leq z_m^{s_j}R_m^{s_j} ~ \forall j, \notag\\
           & C2, C3, C4, C5, C6, C7 \notag
\end{align}

\textbf{Theorem 1:} If the problem $\mathscr{P}_1$ is feasible, the objective function achieves the optimal solution only when the constraint $\ddot{C1}$ is tight, which means $\sum\limits_{u\in\bm U_m^{s_j}}x_u^{m, j}y_u^{m, j}R_u^{m, j} = z_m^{s_j}R_m^{s_j}, \forall j\in\bm S_m$ must hold when the objective function gets the maximum value.

\textit{Proof:} We will prove \text{Theorem 1} by a simple contradiction statement. Assume that the optimal resource allocation scheme (e.g., ${x_u^{m,j}}^*$, ${y_u^{m,j}}^*$, ${z_m^{s_j}}^*$) for problem $\mathscr{P}_1$ is obtained when $\sum\limits_{u\in\bm U_m^{s_j}}{x_u^{m, j}}^*{y_u^{m, j}}^*R_u^{m, j} <{z_m^{s_j}}^*R_m^{s_j}$. Now all SBSs can achieve backhaul data well because the backhaul rate is higher than the access rate. However, there must exist a $\Delta p>0$ such that $\sum\limits_{u\in\bm U_m^{s_j}}{x_u^{m, j}}^*{y_u^{m, j}}^*R_u^{m, j}=({z_m^{s_j}}^*-\Delta p)R_m^{s_j}$. According to the definition of $Q_i^m(\bm X,\bm Y,\bm Z)$, it's obvious that $Q_i^m({x_u^{m, j}}^*, {y_u^{m, j}}^*, {z_m^{s_j}}^*)>Q_i^m({x_u^{m, j}}^*, {y_u^{m, j}}^*, {z_m^{s_j}}^*-\Delta p)$. That's to say, a larger utility for the VRM can be obtained at ${x_u^{m, j}}^*, {y_u^{m, j}}^*, {z_m^{s_j}}^*-\Delta p$, which also satisfies other constraints. Therefore, there is a contradiction between this conclusion and our assumption. In other words, the optimal cell association and resource allocation scheme is not obtained when $\sum\limits_{u\in\bm U_m^{s_j}}{x_u^{m, j}}^*{y_u^{m, j}}^*R_u^{m, j} <{z_m^{s_j}}^*R_m^{s_j}$. As a result, \text{Theorem 1} is proved.

Based on Theorem 1, we can get $z_m^{s_j}=\frac{1}{R_m^{s_j}}\sum\limits_{u\in\bm U_m^{s_j}}x_u^{m,j}y_u^{m,j}R_u^{m,j}$. Namely, we can replace $\bm Z$ in objective function $\mathscr{P}_1$ by $\bm X$ and $\bm Y$, which means that the original problem will be transformed from a three variables problem into a two variables problem as following:
\begin{align}
\mathscr{\bm {P}}_1:~ &\max\limits_{\bm X, \bm Y}~\sum\limits_{i=1}^N\sum\limits_{m=1}^M\sum\limits_{u\in\bm U_i}\delta_u\mathbb{U}(C_u^m(\bm X, \bm Y))\notag\\
&-\sum\limits_{i=1}^N\sum\limits_{m=1}^M\gamma^mT_i^m(\bm X, \bm Y) \notag \\
&-\sum\limits_{m=1}^M\sum\limits_{j\in\bm S_m}\frac{1}{R_m^{s_j}}(1-\alpha_m)P_m(\sum\limits_{u\in\bm U_m^{s_j}}x_u^{m,j}y_u^{m,j}R_u^{m,j})^2 \\
s.t. ~ & C2, C3, C4, C5 \notag \\
           & C6: \frac{1}{R_m^{s_j}}\sum\limits_{u\in\bm U_m^{s_j}}x_u^{m,j}y_u^{m,j}R_u^{m,j}\leq 1 \notag\\
           & C7: \sum\limits_{j\in\bm S_m}\frac{1}{R_m^{s_j}}\sum\limits_{u\in\bm U_m^{s_j}}x_u^{m,j}y_u^{m,j}R_u^{m,j}\leq 1 \notag
\end{align}
Although problem $\mathscr{\bm {P}}_1$ is simplified based on \text{Theorem 1}, the above problem $\mathscr{\bm {P}}_1$ is still difficult to solve based on the following observations:
\begin{itemize}
\item The feasible set of $\mathscr{\bm {P}}_1$ is non-convex as a result of the binary variables $x_u^{m, j}$.
\item The objective function is not convex due to the product relationship between $x_u^{m, j}$ and convex function of $y_u^{m, j}$.
\end{itemize}
As a result, we first transfer this problem into a convex problem and solve it via a distributed ADMM-based algorithm.

\subsubsection{Transferring $\mathscr{\bm {P}}_1$ into convex problem }\label{section 4_A_1}

As is well known, a mixed discrete and non-convex optimization problem is expected to be very challenging to find its global optimum. Thus, we have to simplify problem $\mathscr{\bm {P}}_1$. Following the approach in \cite{QBYA2013}, we relax $x_u^{m, j}$ ($j=0$ is included) in $\mathscr{\bm {P}}_1$ C2 and C3 to be real value variables such that $0\leq x_u^{m, j}\leq 1$. The relaxed $x_u^{m, j}$ can be interpreted as the time sharing factor that represents the ratio of time when user $u$ associates to the $m$-th MBS or SBS $S_m^j$. However, even after relaxing the variables, the problem is still non-convex due to the non-convex objective function. Thus, to make the problem $\mathscr{\bm {P}}_1$ tractable and solvable, a second step is necessary. Next, we give a proposition of the equivalent problem of $\mathscr{\bm {P}}_1$.

\textit{Proposition:} If we define $\tilde{y}_u^{m, j}=y_u^{m, j}x_u^{m, j}, \forall j$ and $x_u^{m, j}\log\frac{\tilde y_u^{m, jR_u^{m, j}}}{x_u^{m, j}}=0$ for $ x_u^{m, j}=0$, there exists an equivalent formulation of $\mathscr{\bm {P}}_1$ as shown in (\ref{eq:p_1'}).
\newcounter{MYtempeqncnt}
\begin{figure*}[!t]
\normalsize
\setcounter{MYtempeqncnt}{\value{equation}}
\setcounter{equation}{23}
\begin{align}
\label{eq:p_1'}
\mathscr{\bm {P}}_1':&~ \max\limits_{\bm X, \tilde{{\bm Y}}}~ \sum\limits_{i=1}^N\sum\limits_{m=1}^M\sum\limits_{u\in\bm U_i}\delta_u\underbrace{\left\{x_u^{m, 0}\log \frac{\tilde{y}_u^{m, 0}R_u^{m, 0}}{x_u^{m, 0}}+\sum\limits_{j\in\bm S_m}x_u^{m, j}\log \frac{\tilde{y}_u^{m, j}R_u^{m, j}}{x_u^{m, j}}\right\}}_{\mathbb{U}({C_u^m}')} \notag\\
&-\sum\limits_{i=1}^N\sum\limits_{m=1}^M\gamma^m\sum\limits_{u\in\bm U_m \& u\in\bm U_i}\underbrace{\left\{\tilde{y}_u^{m,0}(\alpha_mB_m\cdot P_m)+w_m\sum\limits_{j\in\bm S_m}\tilde{y}_u^{m,j}(1-\alpha_m)B_m\cdot P_m^s\right\}}_{{T_i^m}'}\\
&-\sum\limits_{m=1}^M\underbrace{\left\{\sum\limits_{j\in\bm S_m}\frac{1}{R_m^{s_j}}(1-\alpha_m)P_m(\sum\limits_{u\in\bm U_m^{s_j}}\tilde{y}_u^{m, j}R_u^{m,j})^2\right\}}_{{Q^m}'} \notag\\
\textmd{s.t.}\quad & C2': 0\leq x_u^{m, j}\leq 1, \forall j, \quad C3': \sum\limits_{m=1}^M\sum\limits_{j=0\&j\in\bm S_m}x_u^{m, j}=1, C4': 0\leq \tilde{y}_u^{m, j}\leq x_u^{m,j}, \quad C5': \sum\limits_{u\in\bm U}\tilde{y}_u^{m,j}=1, \forall m, j\notag\\
          &C6': \frac{1}{R_m^{s_j}}\sum\limits_{u\in\bm U_m^{s_j}}\tilde{y}_u^{m,0}R_u^{m,j}\leq 1,\quad C7':\sum\limits_{j\in\bm S_m}\frac{1}{R_m^{s_j}}\sum\limits_{u\in\bm U_m^{s_j}}\tilde{y}_u^{m,0}R_u^{m,j}\leq 1.\notag
\end{align}
\setcounter{equation}{\value{MYtempeqncnt}}
\hrulefill
\vspace*{4pt}
\end{figure*}


The relaxed problem $\mathscr{\bm {P}}_1'$ can be recovered by substitution of variable $\tilde{y}_u^{m, j}=x_u^{m, j}y_u^{m, j}$ into problem $\mathscr{\bm {P}}_1$ except $x_u^{m, j}=0$. Due to the loss of definition when $x_u^{m, j}=0$, it is not a one-to-one mapping. However, $x_u^{m, j}=0$ certainly holds because of the optimality. Obviously, BS does not allocate any resource to any user if the user does not associate with the BS. Thus, it becomes a one-to-one mapping when the completed mapping between $\{x_u^{m, j}, y_u^{m, j}\}$ and $\{x_u^{m, j}, \tilde{y}_u^{m, j}\}$ is defined as
\begin{equation}
y_u^{m, j}=\left\{
                 \begin{array}{lcl}
                 \frac{\tilde{y}_u^{m, j}}{x_u^{m, j}}&\text{if} &x_u^{m, j}>0 \\
                 0 & \text{if} &\textmd{otherwise}
                 \end{array}
                 \right..
\end{equation}

Holding the \textit{Proposition} and well-known perspective function in convex optimization theory \cite{boyd2009convex}, we have the following theory that gives the convexity of $\mathscr{\bm {P}}_1'$.

\textbf{Theorem 2:} If problem $\mathscr{\bm {P}}_1'$ is feasible, it is jointly convex with respect to all optimization variables $x_u^{m, j}$ and $\tilde{y}_u^{m, j}$.

\textit{Proof}: 
Since the constraints $C2', C3', C4', C5', C6, C7$ are linear, they are obviously convex. Due to the fact that objective function $\mathscr{\bm {P}}_1'$ is a linear sum of $\mathbb{U}({C_u^m}')$, ${T_i^m}'$ and ${Q_i^m}'$, we just need to prove the convexity of $\mathbb{U}({C_u^m}')$, ${T_i^m}'$ and ${Q_i^m}'$, based on the following principle:  \textit{the linear sum of convex functions is still convex}\cite{boyd2009convex}.

The convexity proof of $\mathbb{U}({C_u^m}')$ is similar to \cite{gortzen2012optimality}, which will be described briefly as follows. Firstly, we prove the continuity of function $f(t, x)=x\log(t/x), t\geq0, x\geq0$ at the point of $x=0$. Let $s=t/x$, then
$f(t, 0)=\lim\limits_{x\rightarrow0}x\log\frac{t}{x}=\lim\limits_{s\rightarrow\infty}\frac{t}{s}\log s=t\lim\limits_{s\rightarrow\infty}\frac{\log s}{s}=0$.
Function $f(t,x)=x\log(t/x), t\geq0, x\geq0$ is the well-known \textit{perspective operation} of logarithmic function, and the perspective function of a convex function is also a convex function based on \cite{boyd2009convex}. Since $x_u^{m, j}\log\frac{\tilde{y}_u^{m, j}}{x_u^{m, j}}$ is th perspective function of $\log\tilde{y}_u^{m, j}$, and the function $\log\tilde{y}_u^{m, j}$ is convex about $\tilde{y}_u^{m, j}$, $x_u^{m, j}\log\frac{\tilde{y}_u^{m, j}}{x_u^{m, j}}$ is convex. What's more, function $\mathbb{U}({C_u^m}')$ is a weighted sum of series of convex function $x_u^{m, j}\log\frac{\tilde{y}_u^{m, j}}{x_u^{m, j}}$, so it is also a convex function. Function ${T_i^m}'$ is linear function of $\tilde{y}_u^{m, j}$, and the convexity of it is obvious. Since the function ${Q_i^m}'$ is a sum of quadratic function about $(\tilde{y}_u^{m, j})^2$, we can easily know it is a concave function by solving the second derivative of $(\tilde{y}_u^{m, j})^2$. With the theory that a negative concave problem is a convex problem, we can conclude that the form of the objective function in $\mathscr{P}_1'$ is a linear sum of convex problems. With the addition that all the constraints are convex, the convexity of the $\mathscr{P}_1'$ is proved.

Since problem $\mathscr{\bm {P}}_1'$ is a convex problem, a lot of methods (e.g., interior point method) can be used to solve it. However, with the increase of the number of BSs and users, the size of problem will be very large. Practically, even with a powerful computing center, the overhead of deliver enough local information (e.g., channel status information (CSI)) to the global center is extremely inefficient. Therefore, for the purpose of implementation, a distributed algorithm running on each InP should be adopted.
\subsubsection{ADMM-based solution algorithm of $\mathscr{\bm {P}}_1$}\label{section 4_A_2}


Due to constraint $C3'$ in $\mathscr{P}_1'$, the problem is not separable with respect to different InPs. To apply ADMM to the resource allocation problem $\mathscr{P}_1'$, this coupling must be handled appropriately. Therefore, we introduce \textit{local} copy $\bm{X}_m^{\mathcal{z}}$ of the related \textit{global} cell association variable $\bm X$ for the $m$-th InP. Roughly speaking, each \textit{local} variable can be interpreted as the InP's opinion about the corresponding \textit{global} variable. Naturally, variables $\bm {\tilde{Y}}_m$  is also \textit{local} variables for InP $m$, since the InP operates without any limits from other InP. For the sake of brevity, let us define vector $\bm{\mathcal{A}}_m=(\bm X_m^{\mathcal{z}},\bm {\tilde{Y}}_m)$ to represent the \textit{local} variables of the $m$-th InP.
Inspired by \cite{boyd2009convex}, to deal with the constraints, we introduce an indicator function $g(\bm{\mathcal{A}}_m)$ such that $g(\bm{\mathcal{A}}_m)=0$ when $\bm{\mathcal{A}_m}\in \Phi$; otherwise, $g(\bm{\mathcal{A}}_m)=+\infty$, where $\Phi$ represents the feasible set of problem $\mathscr{P}_1'$. With these notations above, problem $\mathscr{P}_1'$ of maximizing $\tilde{G}_{VRM}$ on set $\Phi$ is equivalent to
\begin{align}
\mathscr{P}_1'':~ \min\limits_{\bm{\mathcal{A}}_m}~&\{-\tilde{G}_{VRM}^m(\bm{\mathcal{A}}_m)+g(\bm{\mathcal{A}}_m)\} \\
\textmd{s.t.}~ &\bm X^{\mathcal{z}}_m-\bm{X}=\bm 0 \notag
\end{align}
 where
 \begin{align}
 \tilde{G}_{VRM}^m(\bm{\mathcal{A}}_m)&=\sum\limits_{i=1}^N\sum\limits_{u\in\bm U_i}\delta_u\mathbb{U}({C_u^m}'(\bm X^{\mathcal{z}}_m, \bm{\tilde{Y}} )\notag \\
 &-\sum\limits_{i=1}^N\gamma^m{T_i^m}'(\bm X^{\mathcal{z}}_m, \bm{\tilde{Y}} )-{Q^m}'(\bm X^{\mathcal{z}}_m, \bm{\tilde{Y}})
 \end{align}
 It can be seen that the objective function is separable across InPs in the virtualized small networks. However, the \textit{global} cell association variables involved in the consensus constraints couple the problem with respect to the InPs. Therefore, the basic idea to solve problem $\mathscr{P}_1''$ is that each InP only determines its \textit{local} variables based on the local information, and VRM is responsible for achieving consensus between the \textit{local} variables and the \textit{global} variable according to the consensus constraint.

We apply ADMM \cite{eckstein2012augmented} for solving the problem in $\mathscr{P}_1''$ in a distributed way. The machinery of the ADMM applied to $\mathscr{P}_1''$ is initiated by forming an \textit{augmented Lagrangian} with respect to the consensus constraints. The augmented Lagrangian not only includes a set of consensus constraints weighted by Lagrange multipliers (conventional Lagrangian), but it also involves an additional regularized quadratic term: a squared $\|L\|_2$ norm with respect to the consensus constraints.

Let $\bm\lambda_m$ be the Lagrange multipliers set associated with consensus constraints of $\mathscr{P}_1''$. Thus, the augmented Lagrangian for $\mathscr{P}_1''$ can be written as
\begin{align}
\label{lagelangri}
\mathfrak{L}_{\rho}(\bm{\mathcal{A}}_m, \bm X)&=-\tilde{G}_{VRM}^m+g(\bm{\mathcal{A}}_m)+\bm\lambda_m(\bm X_m^{\mathcal{z}}-\bm X)\notag\\
&+\frac{\rho}{2}\sum\limits_{m=1}^M\|\bm X_m^{\mathcal{z}}-\bm X\|_2^2
\end{align}
where $\rho\in\mathbb{R}_{++}$ is a positive constant parameter for adjusting the convergence speed of the ADMM. Note that due to the structured interconnections in the virtualized small cell networks, the augmented Lagrangian is separable with respect to the InPs.

Fundamentally, the augmentation can be seen as a penalty term added to the primal objective function \cite{BPCPE11}. The regularization term facilitates the algorithm to drive the \textit{local} and the related \textit{global} variables into consensus. It's worth emphasizing that the solution of the original optimization problem $\mathcal{P}_1''$ is not affected by adding the quadratic regularization term to the Lagrangian since it vanishes for any set of primal feasible variables. The ADMM method consists of sequential optimization phases over the primal variables followed by the method of multipliers update for the dual variables \cite{BPCPE11}. By applying the ADMM to the problem in $\mathscr{P}_1''$, we first minimize the augmented Lagrangian in (\ref{lagelangri}) over the \textit{local} variables, then over the \textit{global} variables, and finally, perform the dual variable update. Thus, the ADMM method consists of the following steps:
\begin{align}
\label{ADMM_X}
\bm{\bm{\mathcal{A}}_m}^{t+1}=&arg~ \min\limits_{\bm X_m^{\mathcal{z}},\bm Y_m}\left\{-\tilde{G}_{VRM}^m(\bm{\mathcal{A}}_m)+g(\bm{\mathcal{A}}_m)\right.\notag\\
&\left.+[\bm\lambda_m^T]^t({{\bm X^{\mathcal{z}}}}-[\bm X]^t)+\frac{\rho}{2}\|{{\bm X^{\mathcal{z}}}}-[\bm X]^t\|_2^2 \right\}
\end{align}

\begin{align}
\label{ADMM_Z}
\bm{X}^{t+1}= &arg~ \min\limits_{\bm X} \left\{\sum\limits_{m=1}^M[\bm\lambda_m^T]^t([\bm X^{\mathcal{z}}_m]^{t+1}-\bm X)\right.\notag \\
&\left.+\frac{\rho}{2}\sum\limits_{m=1}^M\|[\bm X^{\mathcal{z}}_m]^{t+1}-\bm X\|_2^2\right\}
\end{align}
\begin{align}
\label{ADMM_U}
[\bm\lambda_m]^{t+1}=[\bm\lambda_m]^{t}+\rho([\bm X^{\mathcal{z}}_m]^{t+1}-[\bm X]^{t+1})
\end{align}
where $t$ stands for the iteration index of the ADMM algorithm. The basic idea behind the ADMM iterations is that we first minimize the augmented Lagrangian with respect to \textit{local} variables $(\bm X_m^{\mathcal{z}}, \bm {\tilde{Y}}_m)$ in $\bm{\mathcal{A}}$-update, and then the VRM calculates the \textit{global} variable $\bm X$ based on all \textit{local} variables, finally update the Lagrange multipliers. In the following, we will deal with how to update $\bm{\mathcal{A}}_m$, $\bm X$ and $\bm\lambda$ in each iteration. Moreover, the distributed implementation of each update will be also discussed.

\textit{a)} $\bm{\mathcal{A}}_m$-Update: Interestingly, as mentioned earlier, it can be found that $\bm{\mathcal{A}}_m$-update is separable across each InP. Therefore, $\bm{\mathcal{A}}_m$-update can be decomposed into $M$ subproblems, which can be solved locally at each InP. After dropping off the constant terms in (\ref{ADMM_X}) which do not affect the solution, the subproblem to be solved at InP $m$ can be given as
\begin{align}
\label{ADMM-X_m}
\min\limits_{\bm X_m^{\mathcal{z}}, \bm {\tilde{Y}}_m}~& -\tilde{G}_{VRM}^m(\bm X_m^{\mathcal{z}},\bm {\tilde{Y}}_m)\notag \\
&+\sum\limits_{u\in\bm U_m}\sum\limits_{j=0\&j\in\bm S_m}\{({{x^{\mathcal{z}}}_u^{m,j}}-[x_u^{m, j}]^t)^2+\lambda_u^{m,j}{{x^{\mathcal{z}}}_u^{m,j}}\} \\
&\textmd{s.t.} ~ \bm{\mathcal{A}}_m\in\bm\Phi \notag
\end{align}
Obviously, problem (\ref{ADMM-X_m}) is a convex problem because of the convexity of problem $\mathscr{P}_1'$. For the purpose of fast convergence, steepest descent method \cite{boyd2009convex} will be adopted at each InP to realize $\bm{\mathcal{A}}_m$-update.

Recall that we have relaxed the cell association indicator to be a real value between zero and one instead of a Boolean at the beginning of this subsection, hence we have to recover it to a Boolean after we get the optimal solution of subproblem (\ref{ADMM-X_m}). Assume ${x_u^{m,j}}^*$, ${y_u^{m, j}}^*$ (including $j=0$ ) and ${z_m^{s_j}}^*$ are the optimal solution of (\ref{ADMM-X_m}) obtained directly from the steepest descent method. Inspired by \cite{WCLM99}, we first compute the marginal benefit for each $x_u^{m,j}$ as follows. $D_u^{m,j}=\frac{\partial L_\rho({\bm{\mathcal{A}}}_m)}{\partial x_u^{m,j}}$, where $\partial L_\rho({\bm{\mathcal{X}}}_m)$ is the objective function of problem (\ref{ADMM-X_m}). Then, the indicator ${x_u^{m,j}}^*$ can be recovered to zero or one by
\begin{equation}
\label{E-liear-to-Bool}
{x_u^{m,j}}^*=\left\{
                 \begin{array}{lcl}
                 1&\text{if} &{x_u^{m,j}}^*=\max\limits_{m,j}D_u^{m,j}~ \text{and}~ D_u^{m,j}\geq0 \\
                 0& \text{otherwise} \\
                 \end{array}
                 \right..
\end{equation}
After recovering the indicator ${x_u^{m,j}}^*$, we resolve the problem (\ref{ADMM-X_m}) to get the optimal solution of $\tilde{y}_u^{m, j}$ according to the known recovered ${x_u^{m,j}}^*$, which is easy due to the fact that the problem (\ref{ADMM-X_m}) will be concave respect to $\tilde{y}_u^{m, j}$. Note that this is a common method to deal with the indicator variables in resource allocation, which is widely adopted in the literature (e.g., \cite{WCLM99}).

\textit{ b) $\bm{X}$-Update}: Due to the added quadratic regularization term in the augmented Lagrangian (\ref{lagelangri}), the objective in (\ref{ADMM_Z}) is strictly convex in $x_u^{m,j}$. Therefore, the unique optimal solution is found by setting the derivative to zero that results in
\begin{equation}
\label{global_variable_final}
\bm X=\frac{1}{M}\sum\limits_{m=1}^M[\bm X_m^{\mathcal{z}}]^{t+1}+\frac{1}{M\rho}\sum\limits_{m=1}^M[\bm\lambda_m]^t
\end{equation}
By using $\sum\limits_{m=1}^M[\bm\lambda_m]^t=0$, it will result that $\bm X=\frac{1}{M}\sum\limits_{m=1}^M[\bm X_m^{\mathcal{z}}]^{t+1}$ at each iteration\cite{BPCPE11}. Namely, the \textit{global} variables are obtained at each iteration $t$ by averaging out the corresponding updated \textit{local} copies. This introduces a philosophy of interpreting the current \textit{local} copies as InP's opinions about the optimal \textit{global} variables. Since the Lagrange multipliers are not involved in (\ref{global_variable_final}), the local communication overhead is reduced in this step.

\textit{c) $\bm\lambda$-Update :} Compared to $\bm{\mathcal{A}}_m$-update and $\bm{X}$-update, the process of $\bm\lambda$-update is quite simple. After receiving the updated \textit{global} variables $\bm{\bm{X}}^{t+1}$, $\bm\lambda$-update can be performed directly via (\ref{ADMM_U}) at each InP in each iteration.

The iteration process of ADMM-based solution algorithm is concluded in \textbf{Algorithm 1}.
\begin{algorithm}
\caption{ADMM-based solution algorithm}
\label{Alrorithm 2}
\begin{algorithmic}[1]
\STATE {Initialization} \\
a) At each InP $m$, collect channel state information of all users within its coverage;\\
b) Initialize $\bm{X}^0\in\Phi$, $\bm\lambda^0>\bm0$ and a stop criterion threshold $\xi_2$ at the VRM\\
\STATE {\textbf{while} $\|\bm{\bm{X}}^{t+1}-\bm{X}^t\|>\xi_2$ }\\
a) Broadcast $\bm{X}^t$ and $\bm\lambda^t$ to each InP; \\
b) At each InP $m$, calculate ${\bm{\mathcal{A}}_m}^{t+1}$; \\
c) At the VRM, update $\bm{X}^{t+1}$ by combing the results of ${\bm X_m^{\mathcal{z}}}^{t+1}$ from each InP; \\
d) Update $\bm{X}^{t+1}$ via (\ref{global_variable_final});\\
e) Update $\bm\lambda^{t+1}$ via (\ref{ADMM_U}) at VRM;\\
\textbf{End}\\
\STATE Recover $\bm{Y}$ and $\bm{Z}$
\STATE Output the optimal resource allocation scheme $\bm{X}^*$, $\bm{Y}^*$ and $\bm{Z}*$
\end{algorithmic}
\end{algorithm}

\subsection{Solving $\bm{\alpha}$ under Fixed Resource Allocation Scheme $\bm{X}$, $\bm Y$, and $\bm Z$} \label{section 4_B}
For any fixed feasible resource allocation scheme $\bm{X}$, $\bm{Y}$, and $\bm{Z}$, the original problem $\bm{\mathscr{P}}$ will be reduced to a simplified optimized problem with variables $\bm\alpha$ as follows:
\begin{align}
\mathscr{P}_2:&~\max\limits_{\bm{\alpha}}~\sum\limits_{i=1}^N\left\{\sum\limits_{m=1}^M\sum\limits_{u\in\bm U_i}\delta_u\mathbb{U}(C_u^m(\alpha_m))\right.\notag \\
&\left.-\sum\limits_{m=1}^M\gamma^mT_i^m(\alpha_m)-\sum\limits_{m=1}^MQ_i^m(\alpha_m)\right\} \\
\textmd{s.t.}~ &\dot{C1}: \sum\limits_{u\in\bm U_m^{s_j}}x_u^{m, j}y_u^{m, j}R_u^{m, j}(\alpha_m)\leq z_m^{s_j}R_m^{s_j}(\alpha_m) ~ \forall j ~ \& ~ j\neq0 \notag\\
& C8 \notag
\end{align}
Then we have the following property.

\textbf{Property }: If problem $\mathscr{P}_2$ is feasible, it's jointly convex with respect to the optimization variables $\{\alpha_m, \forall m\}$.

\textit{Proof:} Since the constraints $\dot{C1}$ and $C8$ are linear, they are obviously convex. Nextly, we prove the convexity of the objective function. For convenience, we use $f(\bm\alpha)$ to represent the objective function in $\mathscr{P}_2$, it is easy to know $\frac{\partial^2 f}{\partial{\alpha_m}^2}<0$.
%
%
So, the objective function of $\mathscr{P}_2$ is convex and then the convexity of problem $\mathscr{P}_2$ is proved.

Since problem $\mathscr{P}_2$ is a convex problem, many solutions can be used to solve this problem. In this paper, considering the fast convergence, we still adopt the steepest descent method to solve this problem. What's more, it can be observed that there are no coupled relationship among different $\alpha_m$, so the solution of $\mathscr{P}_2$ can be divided into $m$ subproblems and these subproblems can be solved in each InP.

\subsection{Overall Algorithm: The Distributed Virtual Resource Allocation Algorithm}\label{section 4-C}

Based on the analysis of Subsections \ref{section 4-A} and \ref{section 4_B}, the distributed virtual resource allocation algorithm can be summarized as \textbf{Algorithm 2}. In this algorithm, $\bm{\alpha}$ is a \textit{long time-scale} optimization variable, while $\bm{X}$, $\bm{Y}$ and $\bm{Z}$ are \textit{short time-scale} optimization variables. In a relatively long period, the InPs do not change the values of $\bm\alpha$, and they solve their corresponding subproblems in parallel in each iteration to optimize their local variables by using local CSI information and transmit their local results to the VRM. Actually, each local variable can be interpreted as InP's opinion about the corresponding global variable. Then the VRM collects all the local results and coordinates all the InPs to achieve the global consensus based on the consensus constraint and the regularization function. Upon the convergence of the cell association and resource scheme, the InPs attempt to change the value of $\bm\alpha$ to get the optimal gain, and then the network will start the next round cell association and resource allocation adjustment. By this way, there is no need to exchange CSI information between InPs and VRM, which will reduce the signaling overhead significantly. Furthermore, since the combinatorial nature and non-convexity of considered problem have been removed through the problem transformation in Subsections \ref{section 4-A} and \ref{section 4_B}, the computational complexity to solve the original problem has been reduced to a reasonable level.

\begin{algorithm}
\caption{Distributed virtual resource allocation algorithm of small cell networks with FD self-backhauls and virtualization}
\label{Alrorithm 3}
\begin{algorithmic}[1]
\STATE {Initialization} \\
a) At each InP $m$, collect channel state information of all users within its coverage;\\
b) Initialize $\bm\alpha^0=\bm{0.5}$, $o=0$ and a stop criterion threshold $\xi_1$ at the VRM;\\
\STATE \textbf{while} $\|{G_{VRM}'}^{o+1}-{G_{VRM}'}^o\|_2^2>\xi_1$ \\
\quad Push the value of $\alpha_m^o$ to the $m$-th InP \\
\quad Run \textbf{Algorithm 2}\\
\quad Update $\bm\alpha^o\rightarrow\bm\alpha^{o+1}$ based on the result of $\bm{X}^*$, $\bm{Y}^*$ \\
\textbf{End}
\STATE Output the optimal resource allocation scheme $\bm{X}^{*}$, $\bm{Y}^{*}$, $\bm{Z}^{*}$ and $\bm\alpha^*$
\end{algorithmic}
\end{algorithm}

\section{Simulation Results and Discussions}
\label{Sec 5: simulation}
In this section, the effectiveness of our proposed virtualized small cell networks with FD self-backhauls and distributed virtual resource allocation algorithm will be demonstrated by computer simulations. In the simulations, we consider a $1 Km\times1 Km$ square area covered by two InPs and two MVNOs. In each InP, there are one macro BS and four SBSs. Each MVNO owns some subscribed users. The number of subscribed users in each MVNO will be varied in different simulation scenarios, and they are randomly located in the whole area. The available bandwidth of both of the two InPs are \textit{10 MHz}. The transmit power of the macro BS and the transmit power of the SBS are $46dBm$ and $20dBm$, respectively. The channel propagation model refers to \cite{3gpp.36.211}. One macro BS is located at the center of the left half of the area, while the other macro BS is located at the center of the right half.  The SBSs are randomly deployed in the area. This deployment of BSs is based on the consensus that the location of macro BS is usually calculated by network planning but the location of SBS (e.g., femtocell) may depend on the users.

\subsection{Performance Evaluation of the Proposed Virtualized Small Cell Networks with FD Self-backhauls}
In this subsection, we evaluate the performance of the proposed virtualized small cell networks with self-backhauls by comparing the following schemes: (a) A traditional small cell network without FD self-backhaul and virtualization, which is similar to that in \cite{ye2013user}; (b) a small cell network with virtualization but without FD self-backhaul, which is similar to that in \cite{Zaki2010LTE}; (c) a small cell network with FD self-backhaul but without virtualization, which is similar to that in \cite{SSGBRW13}. Each scheme has the similar system configurations as described above. For the schemes with virtualization, we optimize the total utility function defined in (\ref{G:VRM_final}). For the schemes without virtualization, MVNOs demote to general network operations with one InP, and the utility function is optimized separately. In this subsection, the augmented Lagrangian parameter $\rho$ is set to $5*10^7$, and the small cell discounting price $w$ is set to $10^{-3}$.

\begin{figure}[!t]
\centering
\includegraphics[width=0.45\textwidth]{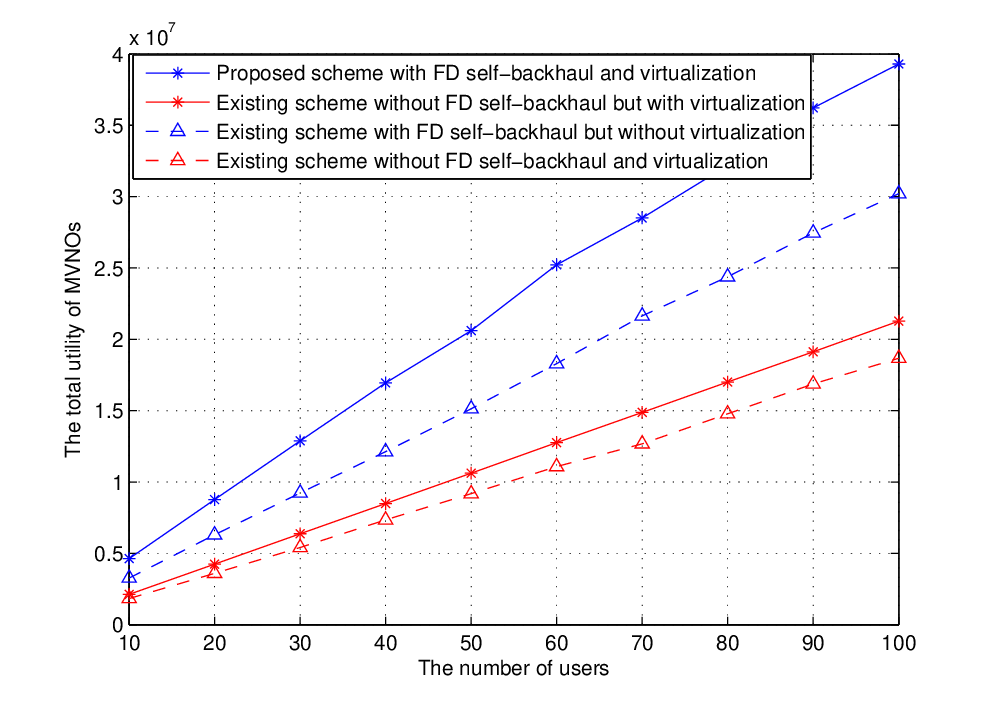}
\caption{The total utility of MVNOs with $\delta_u=10^6$ and $\gamma_i=5$.}
\label{fig:utility of MVNOs}
\end{figure}

\begin{figure}[!t]
\centering
\includegraphics[width=0.45\textwidth]{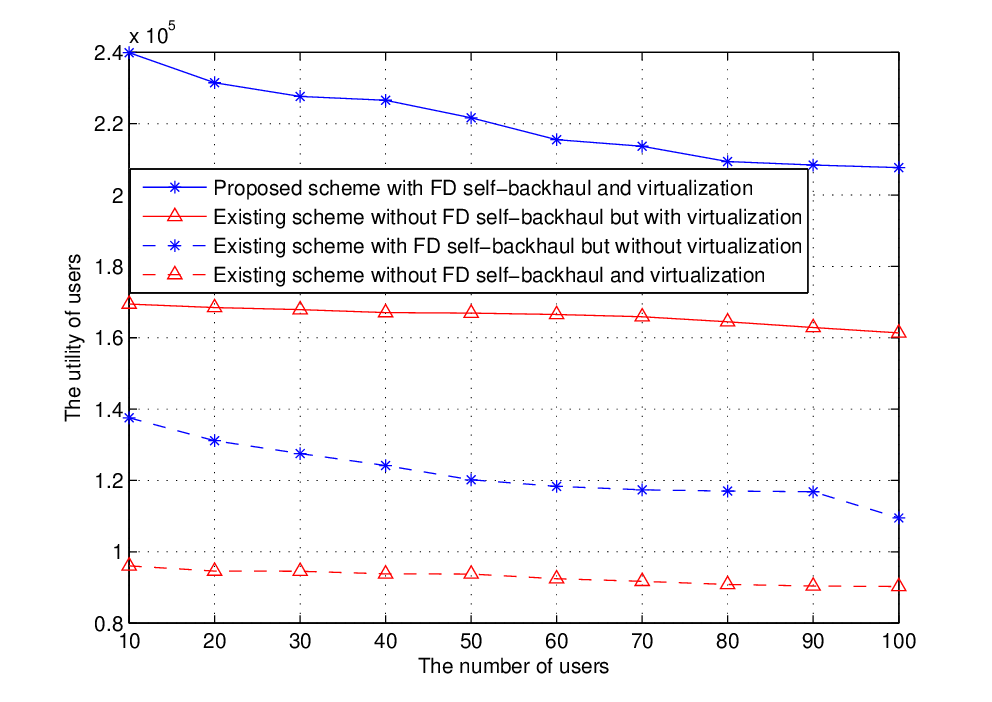}
\caption{The average utility of users with $\delta_u=10^6$ and $\gamma_i=5$.}
\label{fig:utility of users}
\end{figure}

\begin{figure}[!t]
\centering
\includegraphics[width=0.45\textwidth]{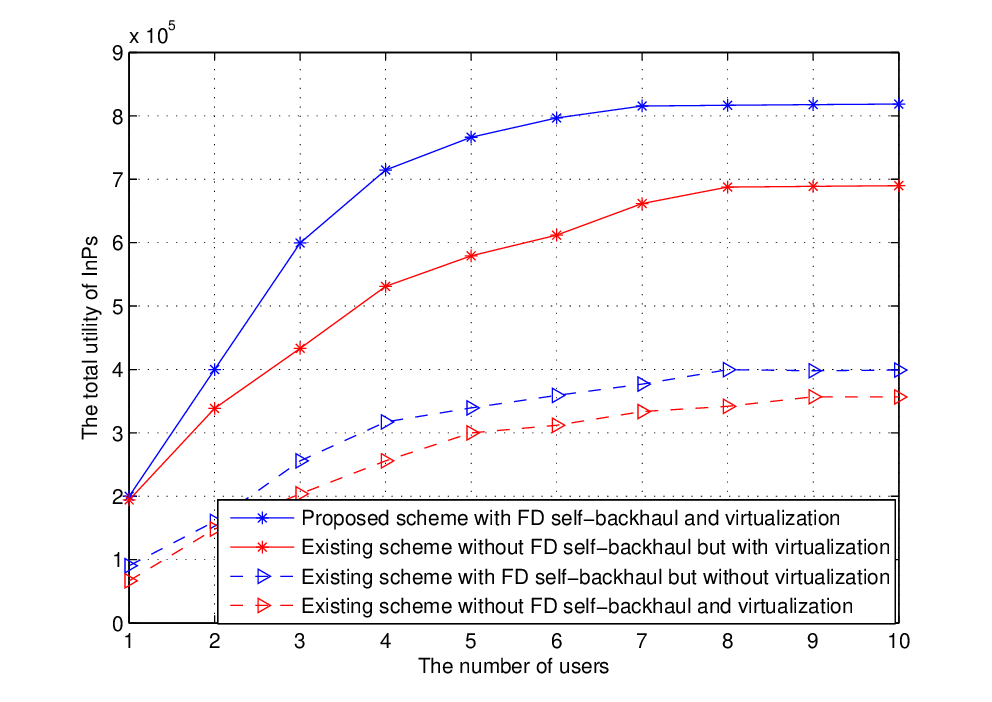}
\caption{The total utility of InPs with $\delta_u=10^6$ and $\gamma_i=5$.}
\label{fig:utility of InPs}
\end{figure}

\begin{figure}[!t]
\centering
\includegraphics[width=0.45\textwidth]{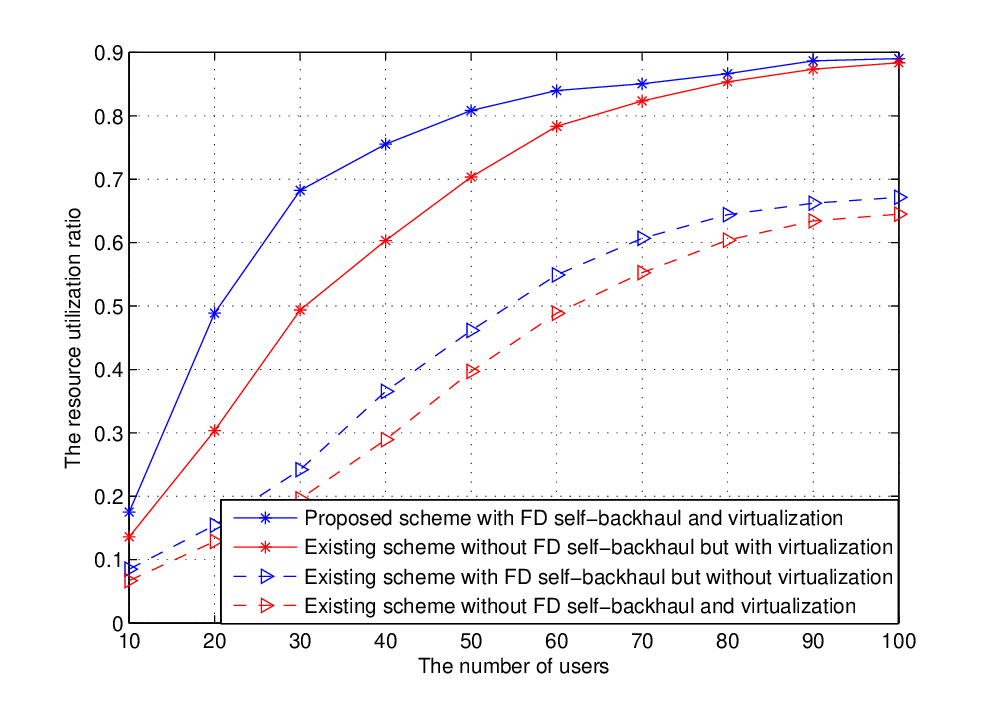}
\caption{The resource utilization ratio with $\delta_u=10^6$ and $\gamma_i=5$.}
\label{fig:resource utilization ratio}
\end{figure}

In Figs. \ref{fig:utility of MVNOs}-\ref{fig:resource utilization ratio}, we compare the utility of MVNOs, users, and InPs, as well as the resource utilization ratio of networks with different schemes. As shown in these four figures, the scheme with FD self-backhaul always outperforms the schemes without FD self-backhaul. This is because the proposed scheme with FD self-backhaul is able to reduce the cost of backhaul and improve the SBS utility ratio, which means that MVNOs can get more available resources at a lower price. As a result, utility values of MVNOs and users are improved. For InPs, the more users access to SBSs, the more backhaul revenue they will get in the small cell network with self-backhaul. However, in traditional small cell networks, the backhaul revenue is used to pay for the backhaul infrastructure rent or construction, and then InPs can only get the resource consumption revenue in access links. This is the reason why the utility of InP with FD self-backhaul is higher than that without FD self-backhaul. Furthermore, there is appreciable performance gain of our proposed scheme compared to traditional schemes without virtualization. The reason is that, with infrastructure virtualization, a user is able to connect to a better access point with better channel conditions and lower resource consumption price. That is to say, access point selection gain and spectrum selection gain can be obtained from our proposed virtualized small cell networks. Therefore, our proposed virtualized  small cell networks with FD self-backhauls can take the advantage of both wireless networks virtualization and FD self-backhauls.

In addition, as shown in Figs. \ref{fig:utility of MVNOs}-\ref{fig:resource utilization ratio}, with the growth of the number of users, the total utility of MVNOs increases linearly, the average utility of users decreases slowly, and the total utility of InPs and the resource utilization ratio of networks will increase, but the increase rate becomes more and more slow. Because MVNOs have to pay money to InPs for using resources, the VRM only allocates optimal resource amount to users rather than all resources. As a result, the total utility of MVNOs will grow since more users will bring more income, and the total utility of InPs and the resource utilization ratio will also go up due to the fact that more resources are consumed. Meanwhile, the average utility of users will descend because some users with bad channel condition access to the network. However, when the number of users is large enough, the ratio of users with bad channel condition will increase and the average link rate of users will decrease accordingly. Considering the resource consumption price, the VRM will not allocate more resources to users because the service rate gain will be lower than the cost of consuming resources. So, the resource utilization ratio and the total utility of InPs will grow no more. Nevertheless, the total utility of MVNOs will keep increasing because of the multi-user diversity gain. This is also the reason why the average utility of users does not decline sharply.

\subsection{The Effect of Self-interference on Our System Performance}
\begin{figure}[!t]
\centering
\includegraphics[width=0.45\textwidth]{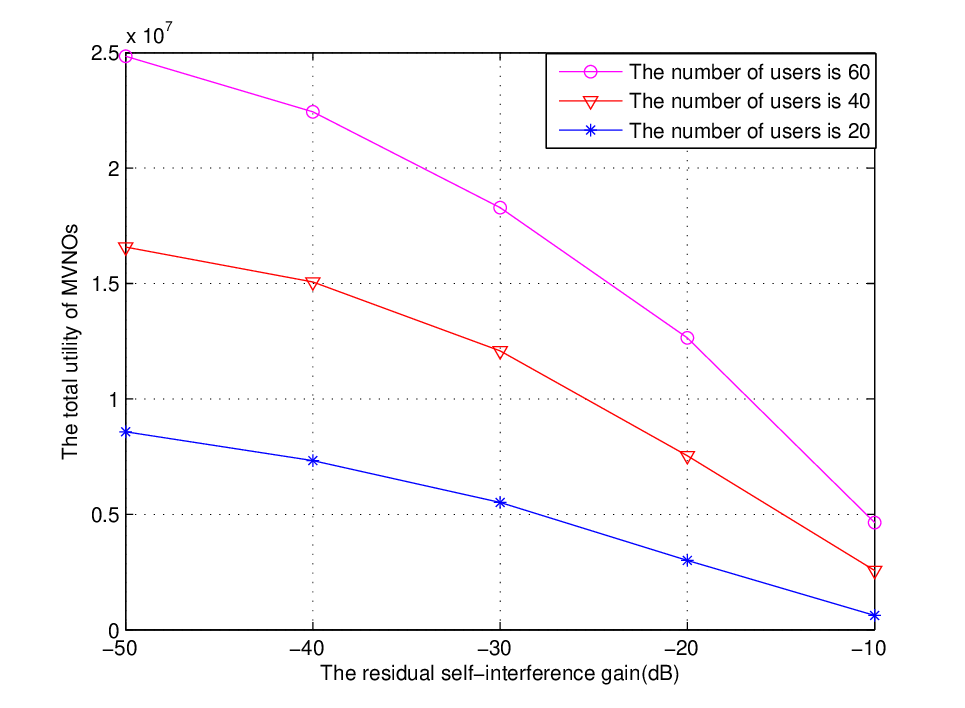}
\caption{The total MVNO utility of different residual self-interference gain with $\delta_u=10^6$ and $\gamma_i=5$.}
\label{fig:self-interference}
\end{figure}

\begin{figure}[!t]
\centering
\includegraphics[width=0.45\textwidth]{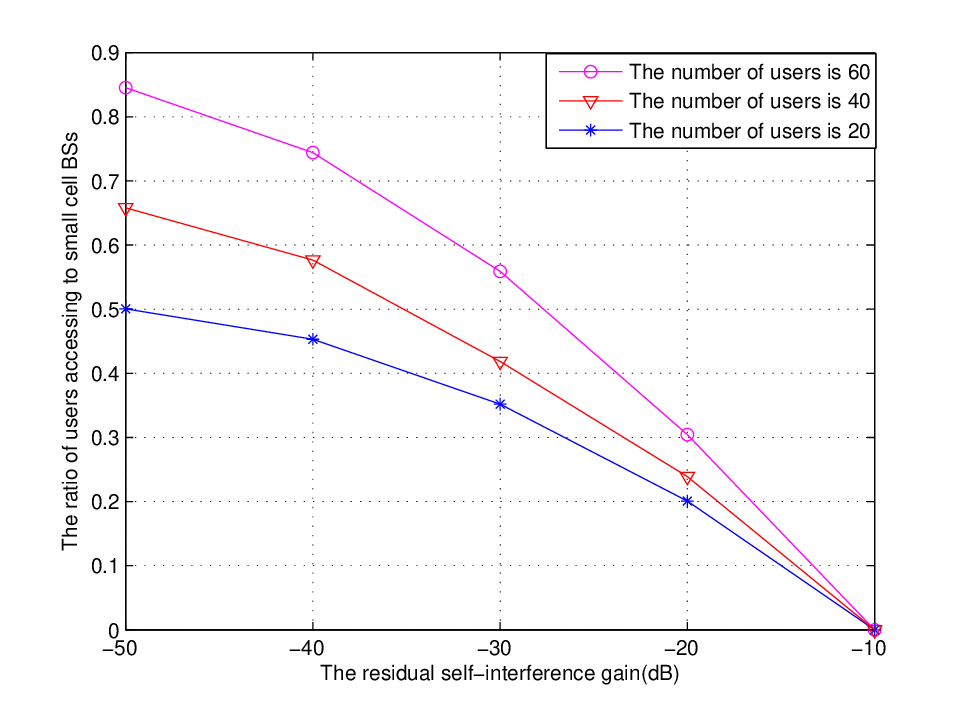}
\caption{The ratio of users accessing to SBSs of different residual self-interference gain with $\delta_u=10^6$ and $\gamma_i=5$.}
\label{fig:self-interference-ratio}
\end{figure}

In Fig.\ref{fig:self-interference} and Fig. \ref{fig:self-interference-ratio}, we evaluate the effect of self-interference on the total utility of MVNOs and the ratio of users accessing to SBSs, respectively. As shown in Fig.\ref{fig:self-interference}, the total MVNO utility will decrease and the rate of decrease speeds up with the increase of the residual self-interference gain. This is because that it will consume more resources for users to access to SBSs because of the terrible backhaul link leading by high residual self-interference gain, and then the users have to access to the relatively expensive MBS, which can be find in Fig. \ref{fig:self-interference-ratio}. What's more, due to the limit of MBSs' resource, the users' service rate getting from MBSs will become lower and lower with the users who access to MBSs increasing. This also is why the total MVNO utility is more sensitive to the residual self-interference gain when there are more users. As shown in Fig. \ref{fig:self-interference-ratio}, there no users accessing to SBSs when the residual self-backhaul gain is $-10$ dB, which waste the resource of SBSs and decrease the spectrum reuse ratio. Therefore, a good self-interference cancellation technology is critical to our FD self-backhaul scheme for SBSs.
\subsection{The Convergence of the Algorithms}
\begin{figure}[!t]
\centering
\includegraphics[width=0.45\textwidth]{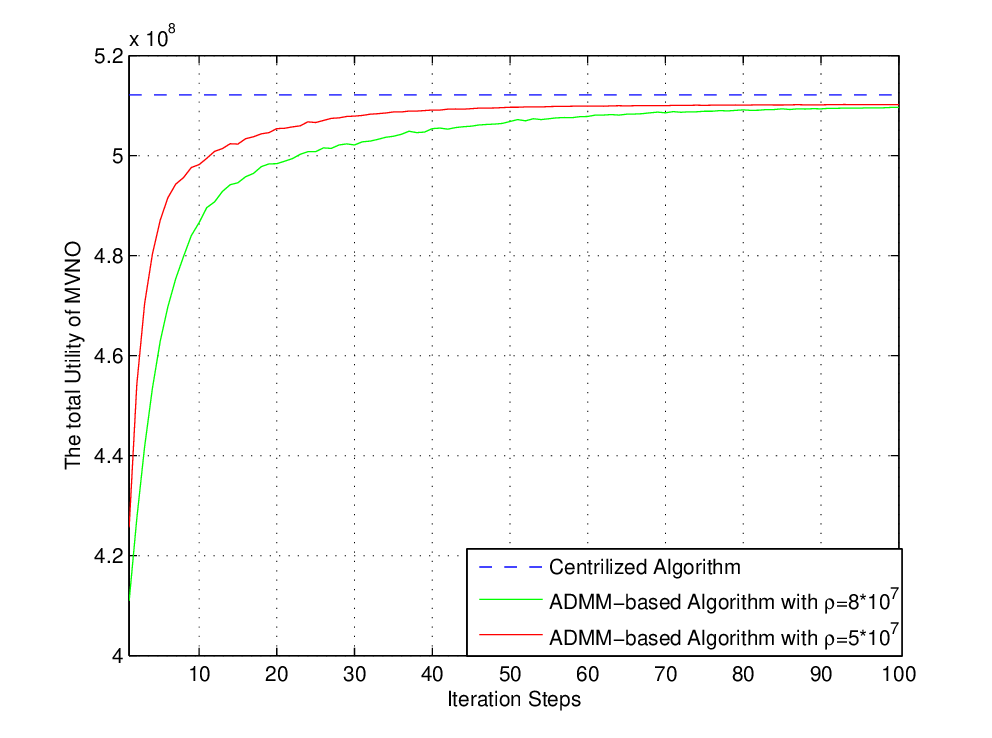}
\caption{The convergence process of ADMM and the effect of parameter $\rho$ ($2*4$ SBSs, $40$ users, $\bm\alpha=[0.5,0.5]$, and $\bm w=[1,1]$). }
\label{fig:convergence process_rho}
\end{figure}

\begin{figure}[!t]
\centering
\includegraphics[width=0.45\textwidth]{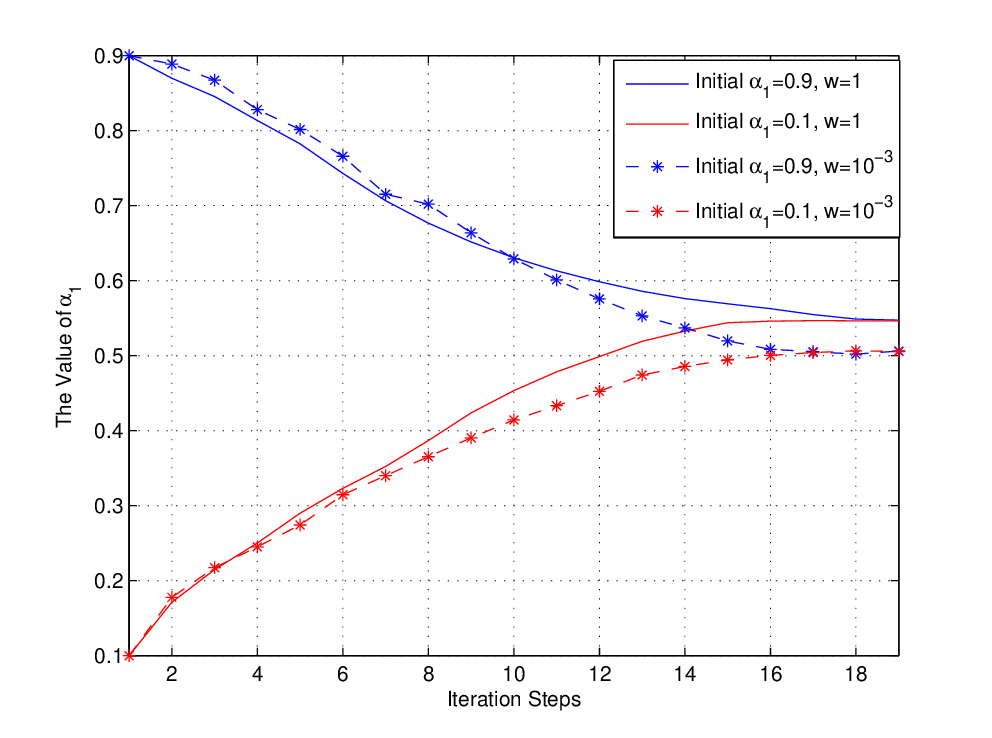}
\caption{The convergence process of $\alpha$ and the effect of parameter $w$ ($2*4$ SBSs, $40$ users, and $\rho=5*10^7$).}
\label{fig:convergence process_different alpha}
\end{figure}

\begin{figure}[!t]
\centering
\includegraphics[width=0.45\textwidth]{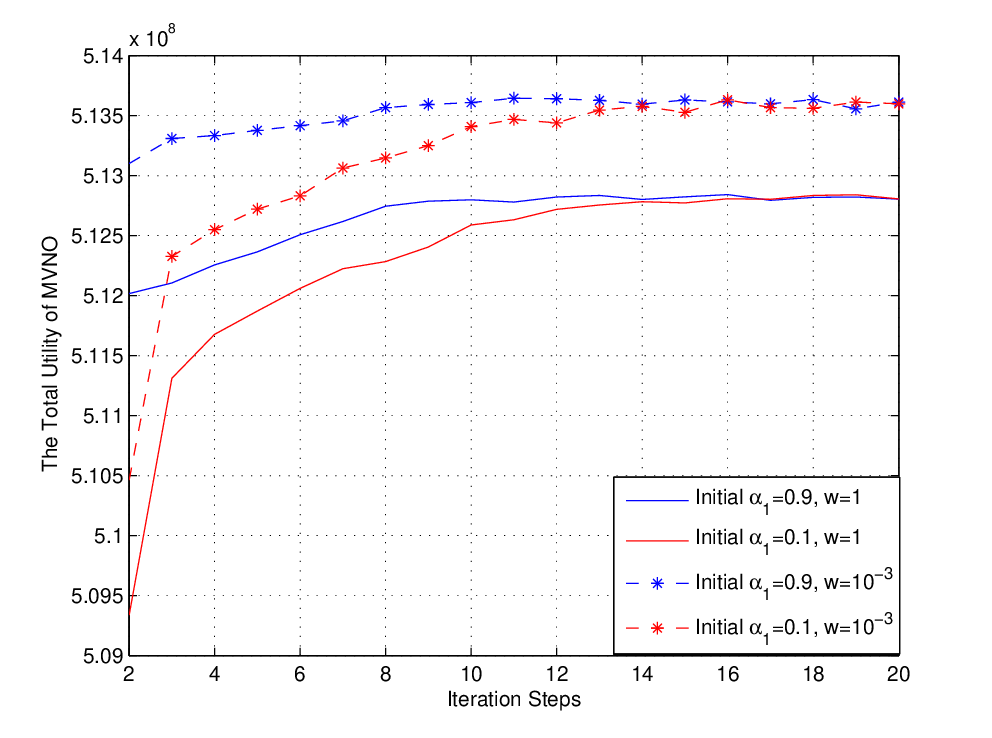}
\caption{The convergence process of overall algorithm and the effect of parameter $w$ ($2*4$ SBSs, $40$ users, and $\rho=10^9$).}
\label{fig:convergence process_w}
\end{figure}
To demonstrate the performance of our proposed scheme further, we show the good convergence performance of our proposed distributed virtual resource allocation algorithm in virtualized small cell networks. The convergence performance includes not only the ADMM-based algorithm in \textbf{Algorithm 1} but also the overall algorithm in \textbf{Algorithm 2}.

Fig. \ref{fig:convergence process_rho} shows the convergence of the proposed ADMM-based \textbf{Algorithm 1} and the effect of parameter $\rho$ in ADMM. As shown in this figure, the gap between the ADMM-based algorithm and the centralized algorithm is narrow, which means the effectiveness of ADMM-based algorithm is equivalent to the centralized algorithm in terms of the overall utility. It can be found that the results with different $\rho$ finally converge to almost the same utility value with  only a small gap. However, the value of $\rho$ affects the rate of convergence. $\rho=5*10^7$ gives higher convergence rate than $\rho=8*10^7$ especially before the $10$-th iteration. Furthermore,  a significant decrease of utility gap between centralized algorithm and proposed ADMM-based algorithm can be found from the $1$-st iteration to the $10$-th iteration. After the $10$-th iteration, the gain of more iterations is still increasing but with less rate. Thus, a tradeoff exits between acceptable utility value and iteration steps. What's more, after multiple experiments, we find the order of magnitudes of $\rho$ must approach that of $\tilde{G}_{VRM}^m$; otherwise, the proposed ADMM-based algorithm will not converge.

Fig. \ref{fig:convergence process_different alpha} and Fig. \ref{fig:convergence process_w} show the convergence of the overall algorithm in \textbf{Algorithm 2} and the influence factors of convergence. As shown in Fig. \ref{fig:convergence process_different alpha}, the final value of $\alpha_1$ almost converges to one constant, although they  begin from different initial values. Similarly, the total utility of MVNOs with different initial $(\alpha 1, \alpha 2)$ are very close in Fig. \ref{fig:convergence process_w}. This shows the robustness  of our proposed overall algorithm with different initial values of $\bm\alpha$. InPs can derive new cell association and resource allocation policies when the network situation changes (e.g., new users arrive), rather than waiting for the related message from VRM, which presents the improvement of self-adaption ability of the network. What's more, it can be found in Fig. \ref{fig:convergence process_different alpha} that the value of $w$ has an influence on the final value of $\alpha$. When $w=1$, it means no discount for MVNO to consume small cell resources, and then the MBS has some superiority to associate more users compared to SBSs because the difference of transmission parameters (e.g., transmission power, antenna gain, the height of BS tower, etc.). However, when $w=10^{-3}$, this superiority of the MBS will be counteracted by the price discount of SBSs, and then some macro users will tend to access to SBSs. This is the reason why the final convergence value of $\alpha_1$ when $w_1=10^{-3}$ is higher than that when $w_1=1$. Corresponding to Fig. \ref{fig:convergence process_different alpha}, the total utility of MVNOs when $w=10^{-3}$ is higher than that when $w=1$. One reason is that the cost of MVNOs for consuming resources decreases as we described in Fig. \ref{fig:convergence process_different alpha}. Another reason is that the load of the network becomes more balanced by adjusting $w$, and then the utilization ratio of SBSs improves, which results in the average throughput improvement of users.

\section{Conclusion and Future Work}
\label{Sec 6: conclusion}
In this paper, we investigated the virtual resource allocation issues in small cell networks with FD self-backhauls and virtualization. We first introduced the idea of wireless network virtualization into small cell networks, and proposed a virtual resource management architecture, where radio spectrum, time slots, MBSs, and SBSs are virtualized as virtual resources. After virtualization, users can access to different InPs to get performance gain. In addition, we proposed to use FD communications for small cell backhauls. Furthermore, we formulated the virtual resource allocation problem as an optimization problem by maximizing the total utility of MVNOs. In order to solve it efficiently, the virtual resource allocation problem is decomposed into two subproblems. In this process, we transferred the first subproblem into a convex problem and solved it by our proposed ADMM-based distributed algorithm, which can reduce the computation complexity and overhead. The second subproblem can be solved by each InP easily because of its convexity and incoherence among InPs. Simulation results showed that the proposed virtualized small cell networks with FD self-backhauls are able to take the advantages of both wireless network virtualization and FD self-backhauls. MVNOs, InPs, and users could benefit from it, and the average throughput of the small cell networks can be improved significantly. In addition, simulation results also demonstrated the effectiveness and good convergence performance of our proposed distributed virtual resource allocation algorithm. Future work is in progress to consider information-centric networking \cite{FYH14} in our proposed framework.

\bibliography{Chen_References}

\begin{IEEEbiography}[{\includegraphics[width=1in,height=1.25in,clip,keepaspectratio]{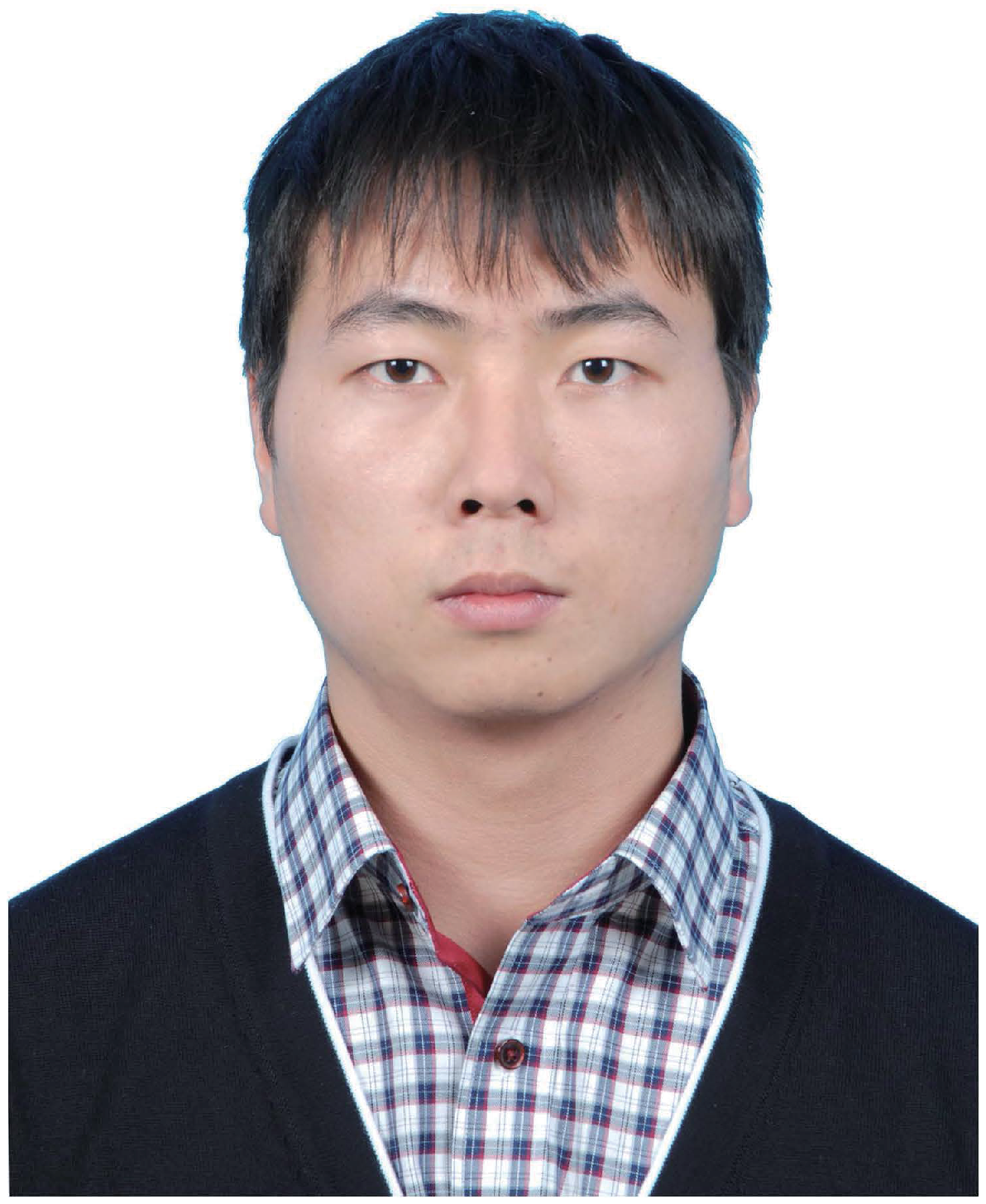}}]{Lei Chen} received the B.S. degree in Communication Engineering from Beijing University of Posts and Telecommunications (BUPT), Beijing, China, in 2011. He is currently working toward the Ph.D. degree with the School of Information and Communication Engineering, BUPT, Beijing, China. He is also with the University of British Columbia as a visiting scholar since Nov. 2014. His current research interests include 5G cellular network, full-duplex wireless, resource management, cross-layer design, and small cell networks.
\end{IEEEbiography}
\vfill

\begin{IEEEbiography}[{\includegraphics[width=1in,height=1.25in,clip,keepaspectratio]{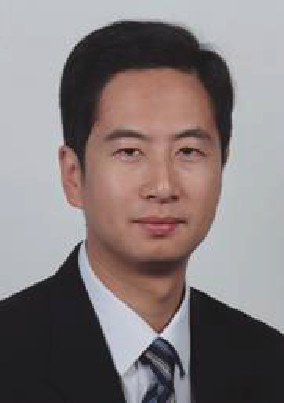}}]{F. Richard Yu} (S'00-M'04-SM'08) received the PhD degree in electrical engineering from the University of British Columbia (UBC) in 2003. From 2002 to 2004, he was with Ericsson (in Lund, Sweden), where he worked on the research and development of wireless mobile systems. From 2005 to 2006, he was with a start-up in California, USA, where he worked on the research and development in the areas of advanced wireless communication technologies and new standards. He joined Carleton School of Information Technology and the Department of Systems and Computer Engineering at Carleton University in 2007, where he is currently an Associate Professor. He received the IEEE Outstanding Leadership Award in 2013, Carleton Research Achievement Award in 2012, the Ontario Early Researcher Award (formerly Premier's Research Excellence Award) in 2011, the Excellent Contribution Award at IEEE/IFIP TrustCom 2010, the Leadership Opportunity Fund Award from Canada Foundation of Innovation in 2009 and the Best Paper Awards at IEEE ICC 2014, Globecom 2012, IEEE/IFIP TrustCom 2009 and Int'l Conference on Networking 2005. His research interests include cross-layer/cross-system design, security, green IT and QoS provisioning in wireless-based systems. 

He serves on the editorial boards of several journals, including Co-Editor-in-Chief for Ad Hoc \& Sensor Wireless Networks, Lead Series Editor for IEEE Transactions on Vehicular Technology, IEEE Communications Surveys \& Tutorials, EURASIP Journal on Wireless Communications Networking, Wiley Journal on Security and Communication Networks, and International Journal of Wireless Communications and Networking, a Guest Editor for IEEE Transactions on Emerging Topics in Computing special issue on Advances in Mobile Cloud Computing, and a Guest Editor for IEEE Systems Journal for the special issue on Smart Grid Communications Systems. He has served on the Technical Program Committee (TPC) of numerous conferences, as the TPC Co-Chair of IEEE GreenCom'15, INFOCOM-MCV'15, Globecom'14, WiVEC'14, INFOCOM-MCC'14, Globecom'13, GreenCom'13, CCNC'13, INFOCOM-CCSES'12, ICC-GCN'12, VTC'12S, Globecom'11, INFOCOM-GCN'11, INFOCOM-CWCN'10, IEEE IWCMC'09, VTC'08F and WiN-ITS'07, as the Publication Chair of ICST QShine'10, and the Co-Chair of ICUMT-CWCN'09. Dr. Yu is a registered Professional Engineer in the province of Ontario, Canada.

\end{IEEEbiography}
\vfill

\begin{IEEEbiography}[{\includegraphics[width=1in,height=1.25in,clip,keepaspectratio]{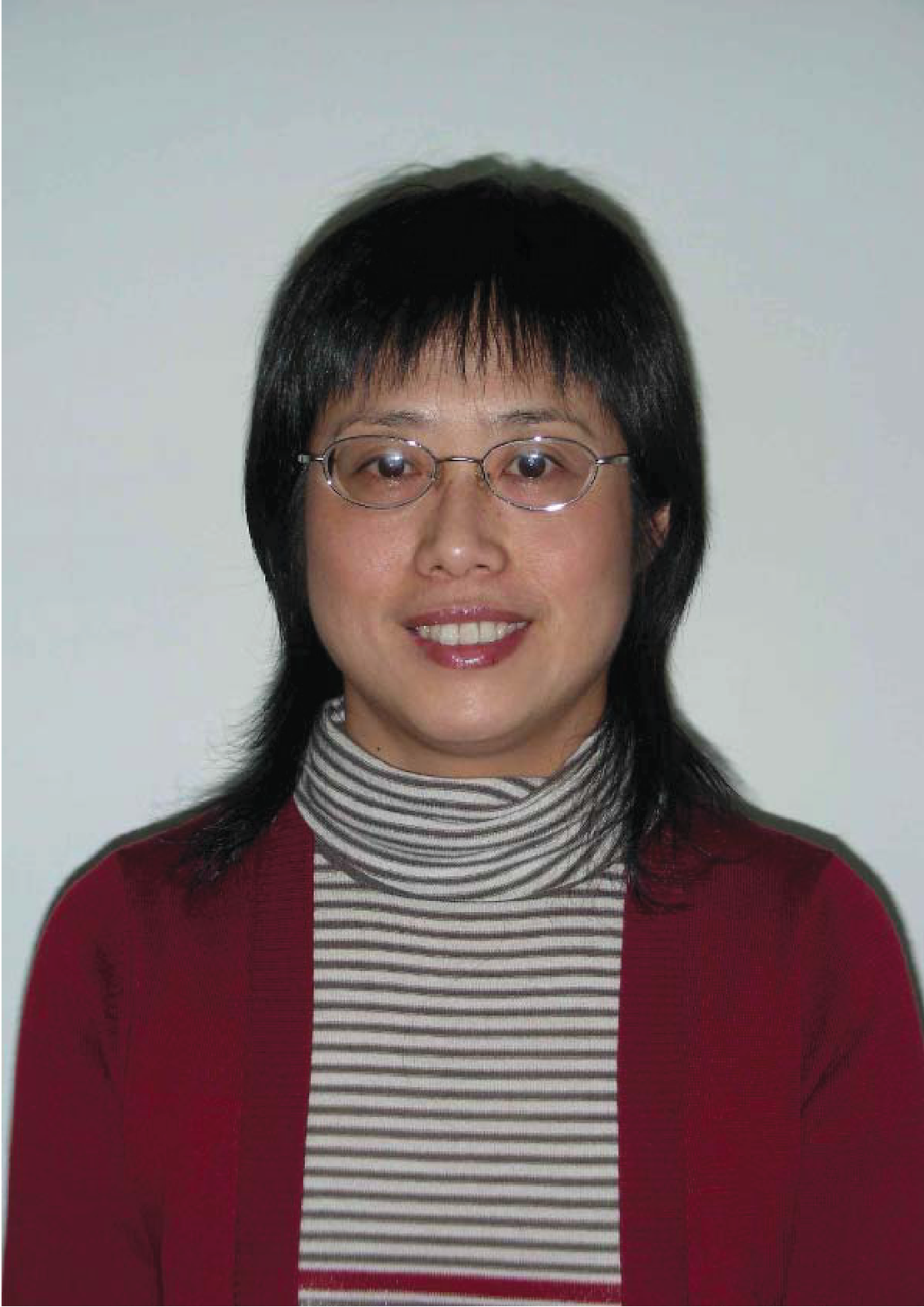}}]{Hong Ji}
received the B.S. degree in communications engineering and the M.S. and Ph. D degrees in information and communications engineering from the Beijing university of Posts and Telecommunications (BUPT), Beijing, China, in 1989, 1992, and 2002, respectively. From June to December 2006, she was a Visiting Scholar with the University of British Columbia, Vancouver, BC, Canada. She is currently a Professor with BUPT. She also works on national science research projects, including the Hi-Tech Research and Development Program of China (863 program), The National Natural Science Foundation of China, etc. Her research interests include heterogeneous networks, peer-to-peer protocols, cognitive radio networks, relay networks, Long-Term Evolution/fifth generation, and cooperative communications.

\end{IEEEbiography}
\vfill

\begin{IEEEbiography}[{\includegraphics[width=1in,height=1.25in,clip,keepaspectratio]{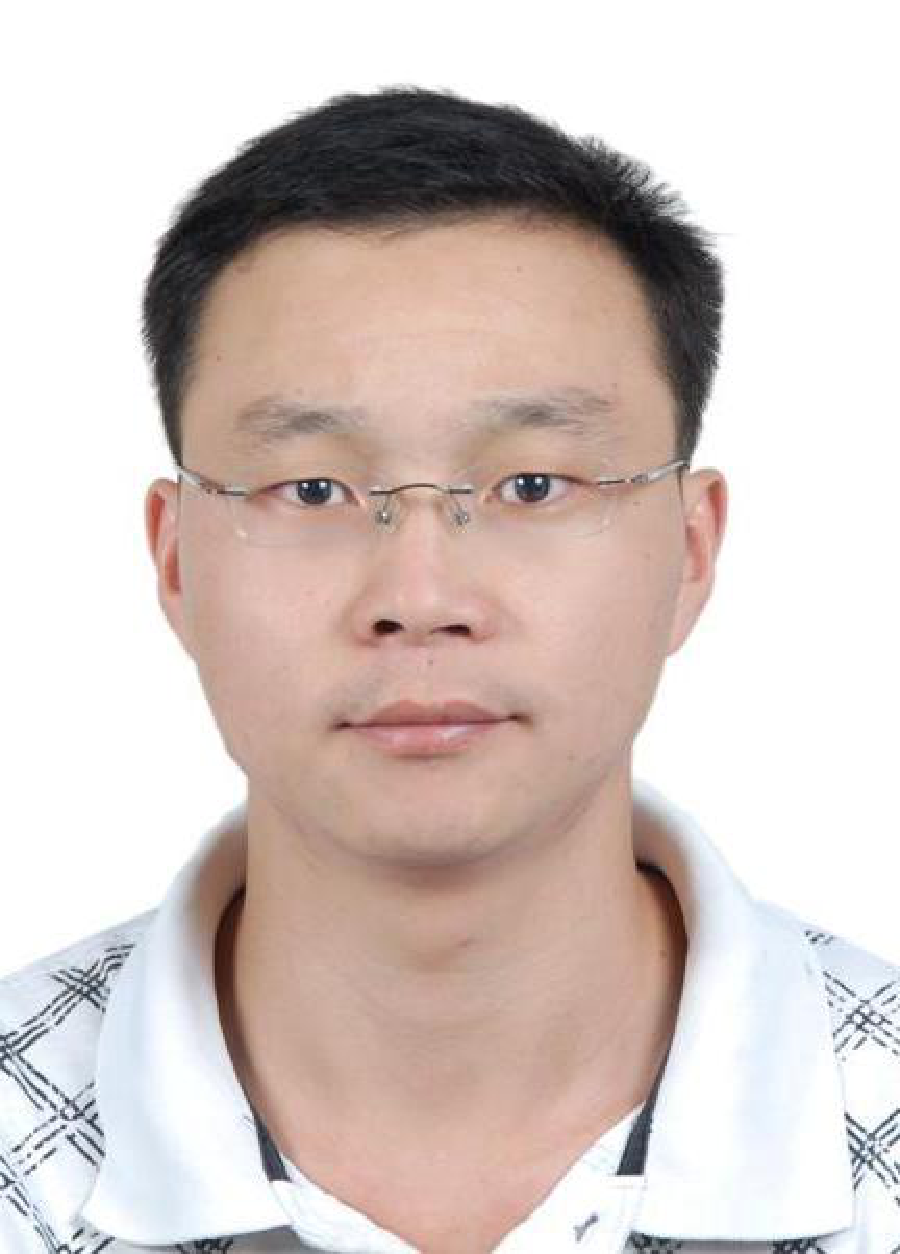}}]{Gang Liu} received the B.S. degree in Electronics and Information Engineering from Sichuan University, Sichuan, China, in 2010. He is currently working toward the Ph.D. degree with the School of Information and Communication Engineering, Beijing University of Posts and Telecommunications, Beijing, China. He is also with the University of British Columbia as a visiting scholar since Nov. 2013. His current research interests include 5G cellular network, full-duplex wireless, resource management, cross-layer design, and TCP/IP protocol in satellite communication systems. He won the second prize in the 2009 National Undergraduate Electronic Design Contest and Best Paper Award in IEEE ICC'2014. He has served as reviewer for International Journal of Communication system, China Communications, ICC'2012, Globecom'2013 and so on.
\end{IEEEbiography}
\vfill

\begin{IEEEbiography}[{\includegraphics[width=1in,height=1.25in,clip,keepaspectratio]{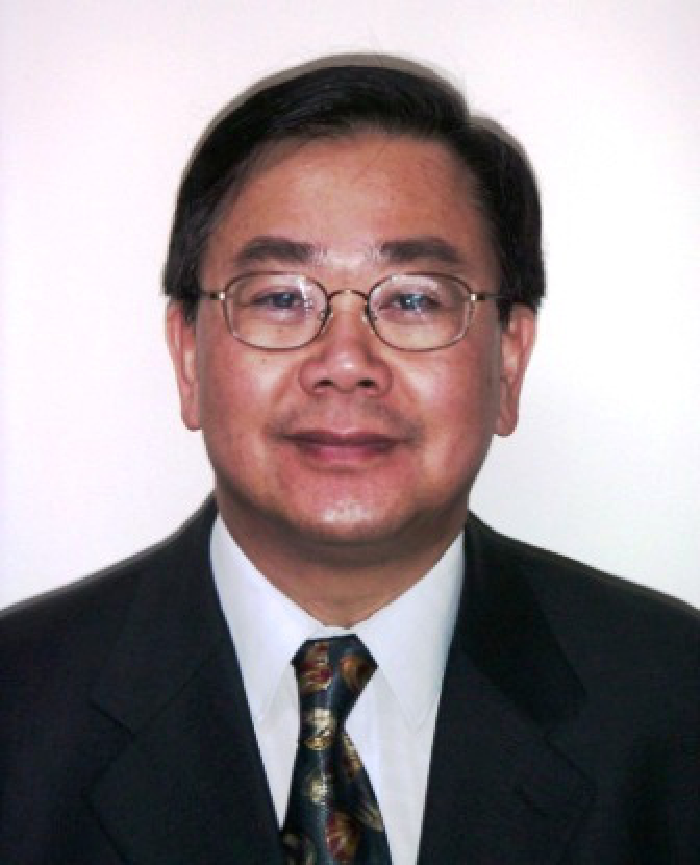}}]{Victor C.M. Leung} (S¡¯75-M¡¯89-SM¡¯97-F¡¯03) received the B.A.Sc. (Hons.) degree in electrical engineering from the University of British Columbia (U.B.C.) in 1977, and was awarded the APEBC Gold Medal as the head of the graduating class in the Faculty of Applied Science. He attended graduate school at U.B.C. on a Natural Sciences and Engineering Research Council Postgraduate Scholarship and completed the Ph.D. degree in electrical engineering in 1981.

From 1981 to 1987, Dr. Leung was a Senior Member of Technical Staffin the satellite systems group at MPR Teltech Ltd. He started his academic career in the Department of Electronics at the Chinese University of Hong Kong in 1988. He returned to U.B.C. as a faculty member in 1989, where he currently holds the positions of Professor and TELUS Mobility Research Chair in Advanced Telecommunications Engineering in the Department of Electrical and Computer Engineering. He is a member of the Institute for Computing, Information and Cognitive Systems at U.B.C. He also holds adjunct/guest faculty appointments at Jilin University, Beijing Jiaotong University, South China University of Technology, the Hong Kong Polytechnic University and Beijing University of Posts and Telecommunications in China. Dr. Leung has co-authored more than 500 technical papers in international journals and conference proceedings, and several of these papers had been selected for best paper awards. His research interests cover broad areas of wireless networks and mobile systems.
\end{IEEEbiography}
\vfill

\end{document}